\documentclass[journal]{IEEEtran}
\usepackage{amsmath,amsfonts}
\usepackage{algorithmic}
\usepackage{bigints}
\usepackage{accents}
\usepackage{hyperref}
\usepackage{algorithm}
\usepackage{array}
\usepackage{subfig}
\usepackage{textcomp}
\usepackage{stfloats}
\usepackage{url}
\usepackage{verbatim}
\usepackage{graphicx}
\usepackage{cite}
\usepackage{amsfonts}
\usepackage{amssymb}
\usepackage{tikz}
\usepackage{mathtools}
\usepackage{circuitikz}
\usetikzlibrary{arrows,
    chains,
    decorations.markings,
    shadows, shapes.arrows,shapes, fit}
\tikzset{%
sum/.style      = {draw, circle, node distance = 2cm}, 
input/.style    = {coordinate}, 
output/.style   = {coordinate}, 
block/.style = { draw,
              thick,
              rectangle,
              minimum height=6.3em,
              fill=none,
              align=center,
},
wide block/.style = {
              block,
              minimum height = 3em,
              text width=2.5cm,
              minimum width = 8em,
},
dotted_block/.style={draw=black!50!white, line width=1pt, dash pattern=on 6pt off 4pt,
            inner sep=4mm, rectangle, rounded corners}
}
\usetikzlibrary{positioning}
\usepackage{pgfplots}
\usepackage{nicefrac}
\usepackage[amssymb]{SIunits}
\usepackage{longtable}

\usepackage{import}
\usepackage{color}
\usepackage{amsthm}
\usepackage{cite}
\usepackage{mathequations}

\pgfplotsset{compat=1.18}
\begin{document}

\title{



{Semi-Supervised Model-Free Bayesian State Estimation from Compressed Measurements } 

}

\author{
    \IEEEauthorblockN{Anubhab Ghosh \IEEEauthorrefmark{1}, Yonina C. Eldar \IEEEauthorrefmark{2},  Saikat Chatterjee \IEEEauthorrefmark{1}} 
    
    \IEEEauthorblockA{\IEEEauthorrefmark{1} Digital Futures Centre and School of Elect. Engg. $\&$ Comp. Sc., KTH Royal Institute of Technology, Sweden} \and \\
    \IEEEauthorblockA{\IEEEauthorrefmark{2} Faculty of Mathematics and Computer Science, The Weizmann Institute of Science, Israel} \\
    \IEEEauthorblockA{\IEEEauthorrefmark{1} anubhabg@kth.se, \IEEEauthorrefmark{2} yonina.eldar@weizmann.ac.il,\IEEEauthorrefmark{1} sach@kth.se}

    \thanks{This work has been submitted to the IEEE for future publication. Copyright may be transferred without notice, after which this version will no longer be available.  
    The work is supported by a research grant from the Digital Futures Centre. Project title: ``DLL: Data-limited learning of complex systems''. Website: https://www.digitalfutures.kth.se/}
}


\maketitle

\begin{abstract}
We consider data-driven Bayesian state estimation from compressed measurements (BSCM) of a model-free process. The dimension of the temporal measurement vector is lower than that of the temporal state vector to be estimated, leading to an under-determined inverse problem. The underlying dynamical model of the state's evolution is unknown for a `model-free process.' Hence, it is difficult to use traditional model-driven methods, for example, Kalman and particle filters. Instead, we consider data-driven methods. We experimentally show that two existing unsupervised learning-based data-driven methods fail to address the BSCM problem in a model-free process. The methods are -- data-driven nonlinear state estimation (DANSE) and deep Markov model (DMM). While DANSE provides good predictive/forecasting performance to model the temporal measurement data as a time series, its unsupervised learning lacks suitable regularization for tackling the BSCM task. We then propose a semi-supervised learning approach and develop a semi-supervised learning-based DANSE method, referred to as SemiDANSE. In SemiDANSE, we use a large amount of unlabelled data along with a limited amount of labelled data, i.e., pairwise measurement-and-state data, which provides the desired regularization. Using {benchmark chaotic dynamical systems}, we {empirically} show that the data-driven SemiDANSE provides competitive state estimation performance for BSCM {using a handful of different measurement systems}, against a hybrid method called KalmanNet and two model-driven methods (extended Kalman filter and unscented Kalman filter) that know the dynamical models exactly.

\end{abstract}


\section{Introduction}
\label{sec:intro}


{Estimating the state of a nonlinear dynamical process from noisy measurements is an active area of research with many applications. Traditionally, a dynamical system is represented using a state space model (SSM). An SSM has two main parts -- a dynamical model and a measurement system. The dynamical model of the process describes the state's evolution over time. The measurement/observation system, illustrates the relation between the unobserved state and the observed, noisy measurement \cite{sarkka2023bayesian, simon2006optimal, barfoot2024state}. 
The Bayesian state estimation task is to find an estimate of the unknown state at a given time instant using available, noisy measurements 
\cite[Chap. 1]{sarkka2023bayesian}.}

{In this article, we assume a compressed, linear measurement system with additive measurement noise, with the measurement system known a prior. The dimension of the measurement vector is smaller than that of the corresponding temporal state vector. The Bayesian state estimation from compressed measurements (BSCM) is an under-determined state estimation problem. Note that BSCM subsumes the special case of Bayesian state estimation from partial measurements where only some components of the state vector are observed.
}

{Classical methods for addressing the BSCM problem are SSM-informed., namely they are \textit{model-driven} and know the under-determined measurement system. Model-driven methods assume knowledge of the dynamical model of the underlying process. Typically, a dynamical model is considered Markovian. Prominent examples of classical methods include the celebrated Kalman filter (KF) and its extensions such as the extended KF (EKF), the unscented KF (UKF) and the sampling-based particle filter (PF) \cite{kalman1961new, kalman1960new, gruber1967approach, julier2004unscented, arulampalam2002tutorial}. KF uses a linear SSM and EKF uses a linearized SSM. UKF and PF, on the other hand, can directly work with the nonlinear SSM. The success of these methods for tackling the BSCM problem is owing to the regularization provided by the dynamical model. 
}

{
Our interest in this article is to address BSCM for a model-free process. Here, `model-free' means that the dynamical model of the process is \textit{unknown}. Therefore, the use of classical model-driven methods is challenging, motivating the need for \textit{data-driven} solutions.} {Design of data-driven methods uses different learning approaches. Learning can be in a supervised, unsupervised or semi-supervised manner. Supervised learning uses a labelled training dataset that consists of pairwise state and noisy measurement trajectories. On the other hand, unsupervised learning considers an unlabelled training dataset comprising noisy measurement trajectories. A semi-supervised learning approach uses a mix of labelled and unlabelled training data -- typically, a limited amount of labelled data along with a larger amount of unlabelled data.
}

{Examples of data-driven methods for Bayesian state estimation are Gaussian process-based (GP-based) methods proposed in \cite{ko2009gp, frigola2013bayesian, svensson2016computationally}. GP-based techniques seek to learn the underlying process dynamics in a data-driven manner, using either a supervised or an unsupervised framework with the help of particle Markov chain Monte Carlo (MCMC). 
A second approach involves learning the unknown process noise or measurement noise statistics using an end-to-end approach via recurrent neural networks (RNNs) \cite{xu2021ekfnet}. {Another strategy involves learning the inverse relation of the measurement system} and performing inference using message passing and graph neural networks \cite{garcia2019combining}. In \cite{jung2020mnemonic}, the Markovian constraint in the process state dynamics is relaxed, and long short-term memory networks (LSTMs) are used for learning the process state dynamics from linear measurements. A semi-supervised approach using differentiable particle filters (DPFs) to explicitly learn the Markovian state dynamics and the measurement system was proposed in \cite{wen2021semidpf}. A class of variational inference-based approaches referred to as dynamic variational autoencoders (DVAEs) that seek to model the underlying process state dynamics using an approximate posterior distribution was proposed in 
\cite{girin2021dynamical, fraccaro2017disentangled, krishnan2015deep, krishnan2017structured}. A relevant example of a DVAE applied to state estimation is the deep Markov model (DMM) proposed in \cite{krishnan2017structured}. Finally, an unsupervised learning-based data-driven nonlinear state estimation (DANSE) method for a model-free process was considered in \cite{ghosh2023danse, ghosh2023dansejrnl}.}

{Hybrid methods combine model-driven approaches with data-driven techniques. One prominent hybrid method is KalmanNet \cite{revach2022unsupervised, revach2022kalmannet} and its varieties \cite{ni2024aknet, choi_split-kalmannet_2023, chen2025maml}. KalmanNet is {usually partially} SSM-informed {(only knowledge of the process and measurement dynamics)} and uses data-driven neural networks. In \cite{li2023target}, a UKF is used together with RNNs to model the process state dynamics using a general non-linear autoregressive moving-average (NARMA) model. We point the interested reader to \cite[Section I-A.]{ghosh2023dansejrnl} for a more comprehensive review of Bayesian state estimation covering model-driven, data-driven and hybrid approaches. 
}

{Both data-driven and hybrid approaches require training data for addressing state estimation tasks. While collecting a sufficient amount of unlabelled training data is realistic (e.g., by observing a model-free process for a long period), collecting a sufficient amount of labelled training data is expensive. Hence, designing a fully supervised, data-driven method for a model-free scenario is challenging. {While unsupervised, data-driven approaches seem feasible, using them for tackling the BSCM problem has limitations}. For example, unsupervised approaches such as DMM \cite{krishnan2015deep, krishnan2017structured} and GP-based SSM \cite{frigola2013bayesian}, are known to be computationally expensive to train and infer due to ancestral sampling. {In the same context, the DANSE method is computationally simple and has been shown to perform well when the Bayesian state estimation problem is not under-determined \cite{ghosh2023dansejrnl}.} {However, when the problem is indeed underdetermined, DANSE suffers in performance and this  limitation was shown using a single experiment with the benchmark chaotic dynamical system -- the Lorenz-$63$ system in \cite[Section III-F]{ghosh2023dansejrnl}.}}

{The experimental evidence leads to the following \emph{hypothesis}: an unsupervised learning-based data-driven method like DANSE fails to tackle the BSCM problem for a model-free process owing to a lack of suitable regularization. Our remark is that the regularization is lacking as DANSE neither has a dynamical model nor access to labelled data. While the evidence of failure is mentioned in  \cite{ghosh2023dansejrnl}, there was no comprehensive study to verify the hypothesis. In this article, we empirically verify this hypothesis for DANSE and another unsupervised learning-based method called DMM. We also address the limitation of unsupervised learning using a \textit{semi-supervised} learning-based approach.} 

{Typically, semi-supervised learning is utilized in scenarios where only a limited amount of labelled data is insufficient to achieve satisfactory performance in machine learning tasks, e.g., classification, regression, clustering, etc. \cite{chapelle2006semi, zhu2009introduction}. While a large body of semi-supervised learning research predominantly considers object detection and image segmentation \cite{yang2022survey, van2020survey}, a comparatively limited number of approaches consider regression tasks \cite{kostopoulos2018semi}. \textit{In this paper, we design a semi-supervised learning-based approach to address the BSCM problem for a model-free process.}
}

{Our main contributions in this work are as follows.  
We develop a semi-supervised learning-based method called semi-supervised DANSE (SemiDANSE) to tackle the BSCM problem for a model-free process. SemiDANSE has similar modeling assumptions as DANSE and retains the same computational benefits at inference time. However, unlike the unsupervised learning-based DANSE, SemiDANSE requires a combination of a limited amount of labelled data along with a large amount of unlabeled data for offline training.  
{Our results are empirical and we do not focus on providing any theoretical guarantees concerning the use of semi-supervised learning in this paper. We experimentally} demonstrate the limitation of unsupervised learning-based methods to handle the BSCM task in the case of two existing approaches -- DANSE and DMM using the benchmark Lorenz-$63$ system \cite{lorenz1963deterministic}. For the Lorenz-$63$ system, we experimentally show the competitive state estimation performance of SemiDANSE compared with the model-driven EKF, UKF, and the hybrid KalmanNet for {a handful of different} under-determined measurement systems. {Finally, we also use three other challenging, chaotic dynamical systems -- a Chen system \cite{vcelikovsky2005generalized, chen1999yet}, a R\"ossler system \cite{rossler1976equation} and a high-dimensional Lorenz-$96$ system \cite{lorenz1996predictability, lorenz1998optimal}} to empirically show that while DANSE fails, SemiDANSE performs reasonably well for the BSCM task. 
}

{The outline of the paper is as follows: in Section \ref{sec:problemformulation}, we restate the BSCM problem mathematically. Next, in Section \ref{sec:semidanse}, we present the proposed SemiDANSE {method}, illustrating the associated inference and learning problems. Then, we perform experiments in Section \ref{sec:experimentsandresults} to show how SemiDANSE can tackle the BSCM problem for a model-free process and compare its performance with relevant model-driven, data-driven and hybrid methods. Finally, we provide our conclusions and {scope of future work} in Section \ref{sec:conclusions}.
}

\subsection{{Notations}}
\label{sec:notations_outline}
We use bold font, lowercase symbols to denote vectors and regular, lowercase font to denote scalars. For example, $\bstate$ represents a vector while $\state_{j}$ represents the $j$'th component of $\bstate$. A sequence of vectors $\bstate_1, \bstate_2, \ldots, \bstate_t$ is compactly denoted by $\bstate_{1:t}$, where $t$ denotes a discrete time index and $\state_{t,j}$ {is} the $j$'th component of $\bstate_t$. Upper case symbols in bold font, like $\bmeasmat$, represent matrices. The operator $\left(\cdot\right)^{\top}$ denotes the transpose. $\normaldist{\cdot}{\stateMean}{\stateCov}$ represents the probability density function of the Gaussian distribution with mean $\stateMean$ and covariance matrix $\stateCov$. 
$\determinant{\cdot}, \E{\cdot} \text{ and } \trace{\cdot}$ {are} the determinant, expectation, and the trace operator, respectively.  The notation $\Vert \bstate \Vert^2_{\mathbf{C}}$ {is} the squared $\ell_2$ norm of $\bstate$ weighted by the matrix $\mathbf{C}$, i.e. $\Vert \bstate \Vert^2_{\mathbf{C}} = \bstate^{\top} \mathbf{C} \bstate$. $\text{diag}(\bstate)$ denotes a square, diagonal matrix with $\bstate$ in its main diagonal. $\cardinality{\Dataset}$ {is} the cardinality of the set $\Dataset$. {A set of $N$ natural indices $\{1,2,\hdots,N \}$ is compactly denoted by $\IndexN{N}$.} $\matrixexp{(\mathbf{A})}$ {is} the matrix exponential of $\mathbf{A}$. $\mathbf{I}_{\statedim \times \statedim}$ denotes the $\statedim \times \statedim$ identity matrix, $\mathbf{O}_{\statedim \times \statedim}$ denotes the matrix of zeros of size $\statedim \times \statedim$.  We mention that our notations are similar to those in \cite{ghosh2023dansejrnl}.     


\section{BSCM Problem and SemiDANSE}\label{sec:semidanse}

In this section, we {describe the proposed SemiDANSE method}. {First, we state the BSCM problem formulation mathematically. Then, we describe the inference problem for SemiDANSE. 
{It is worth mentioning here that the semi-supervised learning mechanism (described later in Section \ref{sec:Semi-supervised learning}) is the primary distinguishing feature between SemiDANSE and DANSE since the inference mechanism for the two methods is the same.} 
Therefore, some of the equations in the following section are similar to those in \cite[Section II-B]{ghosh2023dansejrnl}, which we restate for completeness. } 

\subsection{{BSCM Problem formulation}}\label{sec:problemformulation}
{Consider a nonlinear dynamical process $\{ \bstate_t \}$ generating a sequence of states  $\bstate_1, \bstate_2,\hdots,\bstate_t,\hdots$. The $\statedim$-dimensional state vector $\bstate_t \in \setR^{\statedim}$ is observed using an $\measdim$-dimensional noisy linear measurement vector $\bmeas_t \in \setR^{\measdim}$, s.t.
\begin{equation}
\label{eq:measurement}
\bmeas_t = \bmeasmat \bstate_t + \bmnoise_t, \,\, t=1,2, \hdots,
\end{equation}
where $\bmnoise_t \sim \normaldist{\bmnoise_t}{\boldsymbol{0}}{\bmnoiseCov}$ is Gaussian measurement noise with zero mean and covariance matrix $\bmnoiseCov \in \setR^{\measdim \times \measdim}$. Here $\bmeasmat \in \setR^{\measdim \times \statedim}$ denotes a fixed, measurement matrix. {The measurement system in \eqref{eq:measurement} is assumed to be known, which means $\bmeasmat$ and $\bmnoiseCov$ are known.} The {dynamical model} of the process $\{ \bstate_t \}$ is unknown. The BSCM problem of interest is to find a probabilistic estimate of the state $\bstate_{t}$ given the sequence of noisy, linear measurements $\bmeas_{1:t}$ till the present time instant $t$ (i.e. causally), where $n < m$, leading to compressed measurements.} \\

\subsection{Inference {mechanism of SemiDANSE}}\label{sec:inferenceproblem}

{In this section, we describe the inference mechanism of the proposed SemiDANSE method. The inference procedure involves finding an estimate of $\bstate_{t}$ using the available measurements $\bmeas_{1:t}$. We formulate this by first parameterizing a prior distribution of $\bstate_{t}$ based on a sequence of past measurements $\bmeas_{1:t-1}$. For SemiDANSE, similar to DANSE, we parameterize this prior distribution $p(\bstate_t|\bmeas_{1:t-1})$ as a multivariate Gaussian using an RNN. Specifically, the RNN recursively uses $\bmeas_{1:t-1}$ as the input sequence and provides the parameters of the Gaussian prior -- a mean vector and a covariance matrix.} This is schematically shown in Fig. \ref{fig:rnnGaussianprior}. {The RNN (together with the feed-forward networks) has learnable parameters denoted by $\btheta$}. {The actual RNN that we used is a GRU \cite{choPropertiesNeuralMachine2014}. A brief discussion about our GRU implementation, including its layers, is {provided in Section \ref{sec:trainingtestinghyperparams}} and Appendix \ref{sec:parameterizationGaussianPriorRNN}. 
} 

{Mathematically, we have}
\begin{eqnarray}\label{eq:prior_lik}
\begin{array}{c}
p(\bstate_t \vert \bmeas_{1:t-1}{; \btheta}) \!=\! \normaldist {\bstate_t}{\stateMeanprior{t}(\btheta)}{\stateCovprior{t}(\btheta) }, \\
\text{ {s.t.} } \lbrace \stateMeanprior{t}(\btheta), \stateCovprior{t}(\btheta) \rbrace \triangleq \balpha_{t \vert 1:t-1}(\btheta), \\
\balpha_{t \vert 1:t-1} \left(\btheta\right) = \mathrm{RNN} (\bmeas_{1:t-1}; \btheta), \\
\end{array}
\end{eqnarray}
{with the observation distribution as per \eqref{eq:measurement} being
$p(\bmeas_t \vert \bstate_t ) = \normaldist{\bmeas_t}{\bmeasmat \bstate_t}{\bmnoiseCov}$.}
Here, $\stateMeanprior{t}(\btheta) \in \setR^{\statedim}$ and $\stateCovprior{t}(\btheta) \in \setR^{\statedim \times \statedim}$ denote the mean and covariance matrix of the {parameterized} Gaussian prior respectively. We use $\stateCovprior{t}(\btheta)$ as a diagonal covariance matrix for ease of practical implementation. {In principle, we could have chosen any other parametric distribution for the prior $p\left(\bstate_{t} \vert \bmeas_{1:t-1}; \btheta \right)$, but then we would possibly resort to {a} sampling-based approach using Monte-Carlo approximations for performing the necessary optimization of RNN parameters or subsequent calculation of posterior moments.}

\begin{figure}[t]
    \centering
    \scalebox{0.9}{\begin{tikzpicture}[->,>=latex, auto,node distance=2cm,semithick]

  \node[diamond, minimum size=1.2cm, draw=black, fill=gray!20] (H1) {$\mathbf{z}_{1}$};
  \node[diamond, minimum size=1.2cm, draw=black, right of=H1, fill=gray!20] (H2) {$\mathbf{z}_{2}$};
  \node[diamond, minimum size=1.2cm, draw=black, right of=H2, fill=gray!20] (H3) {$\mathbf{z}_{3}$};
  \node[right of=H3, fill=none] (ellipsisH) {$\ldots$};
  \node[diamond, minimum size=1.2cm, draw=black, right of=ellipsisH, fill=gray!20] (HTminus1) {$\mathbf{z}_{t-1}$};
  
  \node[circle, minimum size=1.2cm, draw=black, below=1cm of H1] (I1) {$\mathbf{y}_1$};
  \node[circle, minimum size=1.2cm, draw=black, right of=I1] (I2) {$\mathbf{y}_2$};
  \node[circle, minimum size=1.2cm, draw=black, right of=I2] (I3) {$\mathbf{y}_3$};
  \node[right of=I3, fill=none] (ellipsisO) {$\ldots$};
  \node[circle, minimum size=1.2cm, draw=black, right of=ellipsisO] (ITminus1) {$\mathbf{y}_{t-1}$};

  \node[rectangle, minimum size=1.2cm, draw=black, above left=1.9cm of HTminus1] (N1) {$\text{FC}_{\mathbf{m}}$};
  \node[circle,fill,inner sep=1.5pt, above=0.4cm of HTminus1] (midnodeHTminus1) {};
  \node[rectangle, minimum size=1.2cm, draw=black, above =1cm of HTminus1] (N2) {$\text{FC}_{\mathbf{L}}$};

  \node[above=0.8cm of N1] (M) {$\mathbf{m}_{t \vert 1:t-1}$};
  \node[above=0.8cm of N2] (L) {$\mathbf{L}_{t \vert 1:t-1}$};
  
   \draw[->, line width=0.75pt] (H1) -- node[right, fill=none, minimum width=4pt, minimum height=4pt] {} (H2);
   \draw[->, line width=0.75pt] (H2) -- node[right, fill=none, minimum width=4pt, minimum height=4pt] {} (H3);
   \draw[->, line width=0.75pt] (H3) -- node[right, fill=none, minimum width=4pt, minimum height=4pt] {} (ellipsisH);
   \draw[->, line width=0.75pt] (ellipsisH) -- node[right, fill=none, minimum width=4pt, minimum height=4pt] {} (HTminus1);

    \draw[->, line width=0.75pt] (I1) -- node[right, fill=none, minimum width=4pt, minimum height=4pt] {} (H1);
    \draw[->, line width=0.75pt] (I2) -- node[right, fill=none, minimum width=4pt, minimum height=4pt] {} (H2);
    \draw[->, line width=0.75pt] (I3) -- node[right, fill=none, minimum width=4pt, minimum height=4pt] {} (H3);
    \draw[->, line width=0.75pt] (ITminus1) -- node[right, fill=none, minimum width=4pt, minimum height=4pt] {} (HTminus1);

   \draw[->, line width=0.75pt] (midnodeHTminus1.west) -| (N1.south);
   \draw[->, line width=0.75pt] (HTminus1.north) -- (N2.south);
   \draw[->, line width=0.75pt] (N1.north) -- (M.south);
   \draw[->, line width=0.75pt] (N2.north) -- (L.south);

   \node[draw, dotted, minimum size=4.5cm, fit=(H1) (H2) (H3) (HTminus1) (ellipsisH) (N1) (N2)] (Border) {};
\end{tikzpicture}}
    \caption{{Schematic of the parameterization of the Gaussian prior $p\left(\bstate_{t} \vert \bmeas_{1:t-1} {; \btheta}\right)$ in \eqref{eq:prior_lik} using an RNN (unfolded in time). The input to the RNN is the sequence $\bmeas_{1:t-1}$ (unshaded, circle nodes). The internal states of the RNN are $\lbrace \rnnhstate_\tau \rbrace$ (shaded, diamond nodes). At the last time instant the hidden state $\rnnhstate_{t-1}$ is mapped to $\stateMeanprior{t}$ and $\stateCovprior{t}$ of the Gaussian prior using shallow, {fully connected}, feed-forward networks with suitable activation functions denoted by $\text{FC}_{\mathbf{m}}$ and $\text{FC}_{\mathbf{L}}$ respectively. {Here, `FC' stands for `Fully Connected.'} The mathematical details of the RNN and the feed-forward networks are described in Appendix \ref{sec:parameterizationGaussianPriorRNN}.}}
    \label{fig:rnnGaussianprior}
\end{figure}
\subsubsection{{Causal state estimation}}\label{sec:causalposteriorestimation}

Using {the prior in \eqref{eq:prior_lik},} the linear measurement setup \eqref{eq:measurement} and the `completing the square' approach \cite[Chap. 2]{bishop2006pattern}, we can compute the posterior $p(\bstate_t|\bmeas_{1:t}{; \btheta})$ in closed-form Gaussian distribution.
\begin{eqnarray}
\label{eq:posterior_update}
\begin{array}{rl}
&p(\bstate_t \vert \bmeas_{1:t}{; \btheta}) =  \normaldist{\bstate_t}{\stateMeanposterior{t}(\btheta)}{\stateCovposterior{t}(\btheta)}, \\
&\stateMeanposterior{t}(\btheta) 
= \stateMeanprior{t}(\btheta) + \mathbf{K}_{t\vert 1:t-1} \pmb{\varepsilon}_t, \\
&\stateCovposterior{t}(\btheta) 
= \stateCovprior{t}(\btheta) -  \mathbf{K}_{t\vert 1:t-1}\mathbf{R}_{\varepsilon, t} \mathbf{K}_{t\vert 1:t-1}^{\top},\\
\end{array}
\end{eqnarray}
where the second equation in \eqref{eq:posterior_update} is obtained using the Woodbury matrix identity/matrix inversion lemma, with 
\begin{equation}\label{eq:posterior_update_additional}
    \begin{array}{l}
    \mathbf{K}_{t\vert 1:t-1}  \triangleq \stateCovprior{t}(\btheta)\bmeasmat^{\top}\mathbf{R}_{\varepsilon, t}^{-1}, \\ \mathbf{R}_{\varepsilon, t}  \triangleq \bmeasmat\stateCovprior{t}\left(\btheta\right)\bmeasmat^{\top} + \bmnoiseCov, \,\, \\
    {\pmb{\varepsilon}}_t  \triangleq \bmeas_t - \bmeasmat\stateMeanprior{t}(\btheta). 
    \end{array}
\end{equation}

{The equations for $p\left(\bstate_t \vert \bmeas_{1:t}; \btheta \right)$ are naturally the same as those for DANSE since the prior distribution in \eqref{eq:prior_lik} is also parameterized as a multivariate Gaussian \cite[Section II-B]{ghosh2023dansejrnl}. A possible point estimate of $\bstate_t$, denoted by $\hat{\bstate}_t\left(\btheta\right)$, can be $\hat{\bstate}_t\left(\btheta\right) = \stateMeanposterior{t}(\btheta)$ {with uncertainty provided by $\stateCovposterior{t}(\btheta)$}}. In that case, we have minimum mean-square-error (MSE) 
{
\begin{eqnarray}
\begin{array}{rl}
     \E{ \| \bstate_t - \hat{\bstate}_t\left(\btheta\right) \|_2^2} &= \E{ \| \bstate_t - \stateMeanposterior{t}(\btheta) \|_2^2} \\
    &= \trace{\stateCovposterior{t}(\btheta)}. 
\end{array}
\end{eqnarray}
}
\subsubsection{{Comparison with inference in KF}}\label{subsec:comparisonwithKFupdates}
{Similar to the KF updates, the linear, Gaussian measurement system in \eqref{eq:measurement} allows for closed-form posterior updates as shown in \eqref{eq:posterior_update} and \eqref{eq:posterior_update_additional} \cite[Chap. 6]{sarkka2023bayesian}. However, we would like to point out 
two crucial differences in the inference steps of SemiDANSE compared to that of traditional KF-based variants.} First, there is no explicit Gaussian propagation from the posterior moments at the current instant to the prior moments at the next time instant, as shown in Fig. \ref{fig:semidanse}. Second, from \eqref{eq:prior_lik}, we note that the prior moments at the $t$'th instant are obtained from the RNN using $\bmeas_{1:t-1}$ as opposed to a dynamical model-driven prediction step using previous posterior moments. 

\subsubsection{{Joint posterior estimation maintaining causality}}\label{sec:jointposteriorestimation}
{An additional task is to find the joint posterior $p\left(\bstate_{1:t} \vert \bmeas_{1:t}{; \btheta}\right)$ because it is a crucial component for semi-supervised learning, shown later in Section \ref{sec:Semi-supervised learning}. We proceed as follows. First, noting that using the product rule, we have the following factorization:
\begin{equation}\label{eq:joint_posterior_product_rule}
    p\left(\bstate_{1:t} \vert \bmeas_{1:t}{; \btheta}\right) = \prod_{\tau=1}^t p\left(\bstate_\tau \vert \bstate_{1:\tau-1} 
, \bmeas_{1:t}{; \btheta}\right).
\end{equation}
In the absence of any simplifying assumptions, the prevalent approach for modelling $p\!\left(\bstate_{1:t} \vert \bmeas_{1:t}{; \btheta}\right)$ using \eqref{eq:joint_posterior_product_rule} involves ancestral sampling of past states $\bstate_{1:\tau-1}$ which is computationally expensive for moderately large $T$ \cite{girin2021dynamical}. We avoid this computational burden by employing a conditional independence assumption -- the current state $\bstate_{\tau}$ is assumed conditionally independent of the previous states $\bstate_{1:\tau-1}$ given the measurements $\bmeas_{1:t}$. We point out that this is often used in variational inference-based approaches for obtaining simplified, mean-field approximations for variational distributions \cite[Chap. 10]{bishop2006pattern}, \cite{krishnan2017structured} and helps further simplify the right-hand side of \eqref{eq:joint_posterior_product_rule}. Then, we have
\begin{equation}\label{eq:posterior_comp_mean_field}
    \prod_{\tau=1}^{t} p\left(\bstate_\tau \vert \bstate_{1:\tau-1} 
, \bmeas_{1:t}{; \btheta}\right) = \prod_{\tau=1}^{t} p\left(\bstate_\tau \vert \bmeas_{1:t}{; \btheta}\right).
\end{equation}
Finally, since we are interested to find an estimate of $\bstate_{\tau}$ using only available measurements $\bmeas_{1:\tau}$, for $\tau = 1, 2, \ldots, t$ (as mentioned in Section \ref{sec:problemformulation}), we use a causality assumption. {Under this assumption, we have} 
\begin{equation}\label{eq:posterior_comp_causality}
\prod_{\tau=1}^{t} p\left(\bstate_\tau \vert \bmeas_{1:t}{; \btheta}\right) = \prod_{\tau=1}^{t} p\left(\bstate_\tau \vert \bmeas_{1:\tau}{; \btheta}\right).
\end{equation}
{The equality mentioned in \eqref{eq:posterior_comp_causality} holds under the causality assumption, which is a modeling assumption in our case}. 
 Combining \eqref{eq:joint_posterior_product_rule}, \eqref{eq:posterior_comp_mean_field} and \eqref{eq:posterior_comp_causality}, we have
\begin{eqnarray}\label{eq:joint_posterior_at_time_t}
\begin{array}{rl}
p\left(\bstate_{1:t} \vert \bmeas_{1:t}{; \btheta}\right) &= \prod_{\tau=1}^t p\left(\bstate_\tau \vert \bmeas_{1:\tau}{; \btheta}\right) \\
&= \, \prod_{\tau=1}^t \! \mathcal{N}(\bstate_{\tau}; \stateMean_{{\tau}|1:{\tau}}(\btheta), \stateCov_{{\tau}|1:{\tau}}(\btheta) ), \\
\end{array}
\end{eqnarray}
where the last equality is obtained by using \eqref{eq:posterior_update}.
}

{Using \eqref{eq:joint_posterior_at_time_t}, we can conveniently incorporate RNNs in a semi-supervised learning framework and maintain causality.} {The use of the assumptions thus provides us with mathematical convenience and entails a trade-off with accuracy in case of modeling the joint posterior \eqref{eq:joint_posterior_at_time_t}.}
In the following {section}, we describe the learning framework of SemiDANSE, which includes learning the unknown parameters $\btheta$ of the RNN.

\section{{Semi-supervised learning of SemiDANSE}}
\label{sec:Semi-supervised learning}

In the following subsections, we formally describe the training dataset for {the learning of SemiDANSE} and our maximum-likelihood based {optimization scheme}. {Finally, we provide intuitions regarding using semi-supervised and unsupervised learning for tackling the BSCM problem.} 

\subsection{{Training data}}\label{sec:training_data}

{Consider} two disjoint subsets of indices as $\IndexSet_s \subseteq \IndexN{N}$ and $\IndexSet_u \subseteq \IndexN{N}$, such that $\IndexSet_s \cup \IndexSet_u = \IndexN{N}$ and $\IndexSet_s \cap \IndexSet_u = \emptyset$. Let us introduce two notations $N_s \triangleq \cardinality{\IndexSet_s}$ and $N_u \triangleq \cardinality{\IndexSet_u}$, such that $N_s + N_u = N$. 

A $T$-length state sequence $\bstate_{1:T}$ has its corresponding measurement sequence $\bmeas_{1:T}$. {Assume} there exists a genie that has access to a dataset 
\begin{eqnarray}
  \Dataset \triangleq \left\lbrace \left(\bstate^{(i)}_{1:T^{(i)}},\bmeas^{(i)}_{1:T^{(i)}}\right)\right\rbrace_{i \in \IndexN{N}},  
\end{eqnarray}
where $T^{(i)}$ is the length of {the} $i$'th sequence. The dataset has $N$ independent state-and-measurement pairwise data samples from the joint random process $p(\{ \bstate_t \}, \{ \bmeas_t \})$. The genie makes a part of the full dataset $\Dataset$ accessible to a designer. When the genie provides the dataset 
\begin{eqnarray}
\label{eq:unsupervised_data}
\Dataset_{u} \triangleq \left\lbrace \left(\bmeas^{(i)}_{1:T^{(i)}}\right)\right\rbrace_{i \in \IndexSet_u} \subset \Dataset
\end{eqnarray}
to the designer, $\Dataset_u$ {consists of} unlabelled (measurement-only)  data for unsupervised learning. If  $\vert\IndexSet_{u}\vert \triangleq N_u = N$, then the designer received the maximum amount of available unlabelled data. On the other hand, if the genie provides the labelled dataset 
\begin{eqnarray}
\Dataset_{s} = \left\lbrace \left(\bstate^{(j)}_{1:T^{(j)}},\bmeas^{(j)}_{1:T^{(j)}}\right)\right\rbrace_{j \in \IndexSet_s} \subseteq \Dataset,
\label{eq:Labelled_Data}
\end{eqnarray}
then, {$\Dataset_{s}$ can be used} for supervised learning. If  $\vert\IndexSet_{s}\vert \triangleq N_s = N$, then the designer received the maximum amount of available labelled data. Finally, if we have access to a dataset $\Dataset_{\text{semi}}$ comprised of two parts as 
\begin{eqnarray}
\label{eq:semisupervised_data}
\Dataset_{\text{semi}} =  \lbrace  \Dataset_{s}, \Dataset_{u} \rbrace,
\end{eqnarray}such that $\cardinality{\IndexSet_{s}} \triangleq N_s \ll N$ and $\cardinality{\IndexSet_{u}} \triangleq N_u = N - N_s$, then $\Dataset_{s}$ is the part of the dataset that is labelled, and $\Dataset_{u} $ is the part of the dataset that is unlabelled. The dataset $\Dataset_{\text{semi}}$ can be used for semi-supervised learning.

\subsection{{Maximum-likelihood based semi-supervised learning}}\label{sec:max_likelihood_semidanse}
{We proceed to describe our maximum-likelihood-based semi-supervised learning approach.} For semi-supervised learning, we start with the joint posterior  
\begin{eqnarray}\label{eq:joint_prob_of_xy}
\begin{array}{rl}
p\!\left(\bstate_{1:t} , \bmeas_{1:t}\right) &\triangleq p\!\left(\bstate_{1:t} , \bmeas_{1:t};\btheta\right) \\
&= p\!\left(\bstate_{1:t} \vert \bmeas_{1:t};\btheta \right) p\!\left(\bmeas_{1:t} ;\btheta \right),
\end{array}
\end{eqnarray}
where the joint distribution of the measurements is as follows:
\begin{eqnarray}\label{eq:prob_of_measurements}
\begin{array}{rl}
p\left(\bmeas_{1:t}\right) \triangleq p\left(\bmeas_{1:t}; \btheta\right) = \prod_{\tau=1}^{t}
p\left( \bmeas_{\tau} | \bmeas_{1:\tau-1} ; \btheta\right). 
\end{array}
\end{eqnarray}
{The conditional marginal distribution $p\left( \bmeas_{\tau} | \bmeas_{1:\tau-1} ; \btheta\right)$} is expressed as in \cite[Section II-B]{ghosh2023dansejrnl}
\begin{eqnarray}
\label{eq:pyt_given_prev}
\begin{array}{l}
p(\bmeas_t | \bmeas_{1:t-1} ) \\
= \displaystyle\int_{\bstate_t} p(\bmeas_t , \bstate_t  |  \bmeas_{1:t-1} ) \, d\bstate_t \\
= \displaystyle\int_{\bstate_t} p(\bmeas_t | \bstate_t )p(\bstate_t |  \bmeas_{1:t-1} ) \, d\bstate_t \\
 = \mathcal{N}(\bmeas_t ; \bmeasmat \stateMeanprior{t}(\btheta), \bmnoiseCov + \bmeasmat \stateCovprior{t}(\btheta) \bmeasmat^{\top}) \\
 \triangleq p(\bmeas_t | \bmeas_{1:t-1}; \btheta ).
\end{array}
\end{eqnarray}
{In the above, we used the prior form \eqref{eq:prior_lik} and the Gaussian observation distribution as per \eqref{eq:measurement}.
Here we introduced the notation $p(\bmeas_{t} \vert \bmeas_{1:t-1}; \btheta )$ to show the explicit dependency of $p( \bmeas_{t} \vert \bmeas_{1:t-1} )$ on $\btheta$.} Note that \eqref{eq:pyt_given_prev} is also a  predictive distribution of $\bmeas_t$ given $\bmeas_{1:t-1}$.



{The} maximum-likelihood based semi-supervised learning problem {using \eqref{eq:semisupervised_data} is} 
\begin{eqnarray}
\label{eq:learning_problem}
\begin{array}{rl}
{\btheta}^{\star} &= \arg\max_{\btheta} \log p(\Dataset_{\text{semi}};\btheta) \\
&= \arg\max_{\btheta} \log p\left(\left\lbrace  \Dataset_{s}, \Dataset_{u} \right\rbrace ;\btheta\right) \\
&= \arg\max_{\btheta} \lbrace \log p\left(\Dataset_{s} ;\btheta\right) + \log p\left(\Dataset_{u} ;\btheta\right) \rbrace. \\
\end{array}
\end{eqnarray}
{We} further expand 
\begin{eqnarray}\label{eq:p_s}
\begin{array}{rl}
&\log p\left(\Dataset_{s} ;\btheta\right) = \log \displaystyle\prod_{j \in \IndexSet_s} p \left(\bstate^{(j)}_{1:T^{(j)}},\bmeas^{(j)}_{1:T^{(j)}};\btheta \right)   \\
&= \log \displaystyle\prod_{j \in \IndexSet_s} p \left(\bstate^{(j)}_{1:T^{(j)}} \vert \bmeas^{(j)}_{1:T^{(j)}} ;\btheta \right) p\left( \bmeas^{(j)}_{1:T^{(j)}} ;\btheta \right),  \\
\end{array}
\end{eqnarray} and 
\begin{eqnarray}\label{eq:p_u}
\begin{array}{rl}
\log p\left(\Dataset_{u} ;\btheta\right) &=   \log \displaystyle\prod_{i \in \IndexSet_u}  p \left( \bmeas_{1:T^{(i)}}^{(i)} ; {\btheta} \right). \\
\end{array}
\end{eqnarray}

Substituting {\eqref{eq:joint_posterior_at_time_t}, 
\eqref{eq:pyt_given_prev},
\eqref{eq:p_s}, \eqref{eq:p_u}} in \eqref{eq:learning_problem}, and converting the logarithm of products to sums, we have the final expression of the semi-supervised learning {optimization problem}: 
\begin{eqnarray}
\label{eq:SemiDANSE_learning}
\begin{array}{rl}
{\btheta}^{\star} &= \arg\max_{\btheta} \Big\lbrace \sum_{j \in \IndexSet_s} \sum_{t=1}^{T^{(j)}} \log \displaystyle p \left(\bstate^{(j)}_{t} | \bmeas^{(j)}_{1:t} ;\btheta \right) \\
& \hspace{0.5cm}+   \sum_{i \in \IndexSet_s \cup \IndexSet_u} \sum_{t=1}^{T^{(i)}} \log p \left( \bmeas_{t}^{(i)} \vert \bmeas_{1: t-1}^{(i)} ; {\btheta} \right) \Big\rbrace \\
&=  \arg\min_{\btheta} \Big\lbrace \sum_{j \in \IndexSet_s} \LossSup\left(\bstate^{(j)}_{1:T^{(j)}}, \bmeas^{(j)}_{1:T^{(j)}}; {\btheta}\right) \\
& \hspace{0.5cm}+ \sum_{i=1}^{N} \LossUnsup\left( \bmeas^{(i)}_{1:T^{(i)}}; {\btheta}\right) \Big\rbrace .  \\
&=  \arg\min_{\btheta} \LossTotal \left(\Dataset_{\text{semi}}; \btheta \right).  \\
\end{array}
\end{eqnarray}
In the above expression, $\LossUnsup\left(\cdot; {\btheta}\right)$ denotes the {loss} due to unsupervised learning, and  $\LossSup\left(\cdot, \cdot; {\btheta}\right)$ is the {loss} due to supervised learning. {Together}, we have a total loss 
\begin{eqnarray}
\begin{array}{rl}
     \LossTotal \left(\Dataset_{\text{semi}}; \btheta \right) &\triangleq \sum_{j \in \IndexSet_s} \LossSup\left(\bstate^{(j)}_{1:T^{(j)}}, \bmeas^{(j)}_{1:T^{(j)}}; {\btheta}\right) \\
    &+ \sum_{i=1}^{N} \LossUnsup\left( \bmeas^{(i)}_{1:T^{(i)}}; {\btheta}\right). 
\end{array}
\end{eqnarray}
\begin{figure}[t]
    \centering
    \scalebox{0.9}{\begin{tikzpicture}[->,>=latex, auto,node distance=2cm,semithick]

\node[rectangle, draw=black, fill=white!50, text width=3.1cm,align=center,rounded corners] (Prior) {$\text{RNN}\left(\cdot;\pmb{\theta}\right)$, \\
$\mathbf{y}_t = \mathbf{H} \mathbf{x}_t + \mathbf{w}_t$,\\ 
$\mathbf{w}_t \sim \mathcal{N}\left(\mathbf{w}_t; \boldsymbol{0}, \mathbf{C}_{{w}}\right)$};
\node [block, fit = (Prior), yshift=1em] (PriorUpdate) {};

\node[wide block, draw=black, fill=white!50, right=2.5cm of PriorUpdate, text width=2cm,align=center] (Posterior) {Posterior Updation};

\node[circle,fill=none,inner sep=0.001pt, right=3.2em of PriorUpdate] (Xdot) {};

\node[rectangle, draw=black, fill=white!50, dotted, below of=Xdot, text width=2cm,align=center] (LossUnsup) { $\mathcal{L}_u\left(\cdot;\pmb{\theta}\right)$};

\node[rectangle, draw=black, fill=white!50, dotted, below=1em of LossUnsup, text width=2cm,align=center] (LossTotal) { $\mathcal{L}\left(\cdot, \cdot;\pmb{\theta}\right)$};

\node[rectangle, draw=black, fill=white!50, dashdotted, below=1em of LossTotal, text width=2cm,align=center] (LossSup) { $\mathcal{L}_s\left(\cdot, \cdot;\pmb{\theta}\right)$};

\node[draw=none, fill=none, below=10.9em of PriorUpdate] (yprev) {$\mathbf{y}_{1:t-1}$};

\node[draw=none, fill=none, right=11.8em of yprev] (ycurrentall) {$\mathbf{y}_{1:t}$};
\node[draw=none, fill=none, right=1em of ycurrentall] (DatasetSemi) {$\mathcal{D}_{\text{semi}}$};
\node[draw=none, fill=none, above=1em of DatasetSemi] (xcurrentall) {$\mathbf{x}_{t}$};
\node[draw=none, fill=none, above=10em of ycurrentall] (ycurrent) {$\mathbf{y}_{t}$};


\node[draw=none, fill=none, above=5em of xcurrentall,  text width=1.5cm,align=center] (posteriorparams) {$\mathbf{m}_{t \vert 1:t}$, \\ $\mathbf{L}_{t \vert 1:t}$};

\node at (PriorUpdate.north) [below, inner sep=2mm] {Prior Updation};

\draw[->, line width=0.75pt] (PriorUpdate.east) -- (Posterior.west) node[draw=none, above=0.1em, text width=1.5cm, align=center, midway]{$\mathbf{m}_{t \vert 1:t - 1}$, \\$\mathbf{L}_{t|1:t-1}$};
\draw[->, dotted, line width=0.75pt] (PriorUpdate.east) -| (LossUnsup.north);

\draw[->, line width=0.75pt] (ycurrentall) -- (yprev.east);
\draw[->, line width=0.75pt] (yprev) -- (PriorUpdate.south);
\draw[->, line width=0.75pt] (ycurrentall) -- (ycurrent);
\draw[->, dotted, line width=0.75pt] (ycurrentall) |- (LossUnsup.east);
\draw[->, line width=0.75pt] (ycurrent) -- (Posterior.222); 
\draw[->, dotted, line width=0.75pt] (DatasetSemi) -- (xcurrentall);
\draw[->,  line width=0.75pt] (DatasetSemi) -- (ycurrentall);
\draw[->, dashdotted, line width=0.75pt] (xcurrentall.north) |- (LossSup.355);
\draw[->, dashdotted, line width=0.75pt] (posteriorparams) |- (LossSup.5);
\draw[->, line width=0.75pt] (Posterior.320) -- (posteriorparams);
\draw[->, dotted, line width=0.75pt] (LossTotal.west) -| (PriorUpdate.300);
\draw[->, dotted, line width=0.75pt] (LossUnsup.south) -- (LossTotal.north); 
\draw[->, dotted, line width=0.75pt] (LossSup.north) -- (LossTotal.south);
\end{tikzpicture}}
    \caption{Schematic of SemiDANSE at the $t$'th time instant. The dotted lines represent information flow specifically during the learning phase, i.e. calculation of losses and gradients for RNN learning, $\Dataset_{\text{semi}}$ represents the training dataset for SemiDANSE as defined in \eqref{eq:semisupervised_data}, $\mathcal{L}_s$ and $\mathcal{L}_u$ denote the supervised and the unsupervised loss respectively at the $t$'th instant as defined in \eqref{eq:supervised_loss} and \eqref{eq:unsupervised_loss} respectively. The dash-dotted lines denote information flow specifically involving the limited $\mathcal{D}_{s}$ in $\Dataset_{\text{semi}}$ during the learning phase. Solid lines represent information flow involving other quantities and during inference.}
    \label{fig:semidanse}
\end{figure}
We note that in the absence of $\LossSup\left(\cdot, \cdot; {\btheta}\right)$, the total loss is only due to unsupervised learning, which is indeed the case for DANSE \cite{ghosh2023dansejrnl}.
We can further expand $\LossSup$ as 
\begin{eqnarray}\label{eq:supervised_loss}
\begin{array}{l}
\LossSup\left(\bstate^{(j)}_{1:T^{(j)}}, \bmeas^{(j)}_{1:T^{(j)}}; {\btheta}\right) \\ \triangleq -\sum_{t=1}^{T^{(j)}} \log \displaystyle p \left(\bstate^{(j)}_{t} \vert \bmeas^{(j)}_{1:t} ;\btheta \!\right)  \\
= \sum_{t=1}^{T^{(j)}} \Big \lbrace  \dfrac{m}{2}\log 2 \pi + \dfrac{1}{2}\log \text{det} \left( \stateCov^{^{(j)}}_{t|1:t}(\btheta) \right)  \\
\hspace{0.5cm} +  \dfrac{1}{2} \| \bstate_t^{(j)} - \stateMeanposterior{t}^{(j)}(\btheta)\|_{\left(\stateCov^{^{(j)}}_{t|1:t}(\btheta)\right)^{-1}}^2 \Big\rbrace, \\
\end{array}  
\end{eqnarray}
where $\stateMeanposterior{t}^{(j)}(\btheta)$ and $\stateCovposterior{t}^{(j)}(\btheta)$ are computed using \eqref{eq:prior_lik}, \eqref{eq:posterior_update}, \eqref{eq:posterior_update_additional} {with input $\bmeas_{1:t}^{(j)}$ where $j \in \IndexSet_s$. This means that the sequence of past measurements $\bmeas_{1:t-1}^{(j)}$ is first passed to the $\text{RNN}\left(\cdot; \btheta\right)$ to obtain the prior moments $\stateMeanprior{t}^{(j)}(\btheta)$, $\stateCovprior{t}^{(j)}(\btheta)$. Subsequently the prior moments, together with the measurement system in \eqref{eq:measurement} and the current measurement $\bmeas_{t}^{(j)}$ are used to compute the posterior moments $\stateMeanposterior{t}^{(j)}(\btheta)$ and $\stateCovposterior{t}^{(j)}(\btheta)$ according to \eqref{eq:posterior_update}, \eqref{eq:posterior_update_additional} where $j \in \IndexSet_s$.} Similarly, we can expand $\LossUnsup$ as follows
\begin{eqnarray}
\label{eq:unsupervised_loss}
\begin{array}{l}
\LossUnsup\left( \bmeas^{(i)}_{1:T^{(i)}}; {\btheta}\right) \\
\triangleq -\sum_{t=1}^{T^{(i)}} \log \displaystyle p \left(\bmeas^{(i)}_{t} \vert \bmeas^{(i)}_{1:t-1} ;\btheta \right)  \\
= \sum_{t=1}^{T^{(i)}} \! \Big \lbrace \! \dfrac{n}{2} \!\log 2 \pi 
\! + \! \dfrac{1}{2}\! \log \text{det} \! \left(\!\bmnoiseCov \! + \! \bmeasmat \stateCovprior{t}^{(i)}(\btheta) \bmeasmat^{\top} \!\right) \\
\hspace{0.5cm} + \dfrac{1}{2} \Vert \bmeas_t^{(i)} - \bmeasmat \stateMeanprior{t}^{(i)}(\btheta)\Vert_{\left(\bmnoiseCov + \bmeasmat \stateCovprior{t}^{(i)}(\btheta) \bmeasmat^{\top}\!\right)^{-1}}^2 \!\! \Big\rbrace. \\
\end{array}  
\end{eqnarray}
{The prior moments  $\stateMeanprior{t}^{(i)}(\btheta)$, $\stateCovprior{t}^{(i)}(\btheta)$ in \eqref{eq:unsupervised_loss} are obtained by passing the sequence of past measurements $\bmeas_{1:t-1}^{(i)}$ to the $\text{RNN}\left(\cdot; \btheta\right)$ for $i \in \IndexSet_u \cup \IndexSet_s$.}

The overall optimization problem \eqref{eq:SemiDANSE_learning}
is non-convex and we solve {it} using gradient descent. {Finally, it is also worth noting that the two loss functions in \eqref{eq:supervised_loss} and \eqref{eq:unsupervised_loss} can be used with the same $\text{RNN}\left(\cdot; \btheta\right)$. To see this, note that \eqref{eq:unsupervised_loss} requires $\stateMeanprior{t}^{(i)}(\btheta)$, $\stateCovprior{t}^{(i)}(\btheta)$ for $i \in \IndexSet_u \cup \IndexSet_s$. Once the prior moments for $i \in \IndexSet_u \cup \IndexSet_s$ are computed, the specific prior moments and the current measurements for the indices $j \in \IndexSet_s$ can be extracted and passed on for additionally computing the posterior moments required in the supervised loss term \eqref{eq:supervised_loss}. {A schematic diagram is depicted in Fig. \ref{fig:semidanse} for visualizing the flow of information during inference and learning. }}

{Finally, we provide some comments on the loss terms in \eqref{eq:learning_problem}.} From \eqref{eq:SemiDANSE_learning}, note that the semi-supervised learning for SemiDANSE translates to the unsupervised learning of DANSE with the maximum amount of unlabelled data, when $\Dataset_s = \emptyset$, that means $\cardinality{\IndexSet_u} \triangleq N_u = N$. A small $N_s \triangleq \cardinality{\IndexSet_s} \ll N$ with $\Dataset_s \neq \emptyset$ is intended and interesting for semi-supervised learning. {Also,} while $\LossSup(\cdot, \cdot; \btheta)$ in \eqref{eq:supervised_loss} has (coincidentally) the same form as that of the loss term for deriving an empirical estimation performance limit in \cite[Section II-E]{ghosh2023dansejrnl}, we reiterate that the purpose of this loss term {here} is completely different. {Lastly, the linear measurement system in enables us to compute the marginal likelihood and the posterior distribution analytically -- leading to direct optimization instead of indirectly maximizing a lower bound of the log-likelihood, as is commonly done in the case of using an expectation lower bound (ELBO) \cite{gordon2017bayesian}.}

\subsection{{Intuitions regarding unsupervised and semi-supervised learning-based methods}}\label{sec:openquestions}
We provide {intuitions} behind the failure of unsupervised learning (DANSE) for addressing the BSCM problem and the success of semi-supervised learning (SemiDANSE) for the same. We reiterate that we do not focus on providing theoretical guarantees concerning supervised or semi-supervised learning, and our supporting results (as shown later in Section \ref{sec:experimentsandresults}) are empirical. {To the best of the authors' knowledge, explainability and performance guarantees for methods involving deep architectures, such as LSTMs and GRUs, is non-trivial and an ongoing research problem in the theoretical machine learning community.}
{The unsupervised learning problem in DANSE \cite{ghosh2023dansejrnl} is }
\begin{eqnarray}
\label{eq:DANSE_learning}
\begin{array}{rl}
\arg\min_{\btheta} \sum_{i=1}^{N} \LossUnsup\left( \bmeas^{(i)}_{1:T^{(i)}}; {\btheta}\right), \\
\end{array}
\end{eqnarray}
where $\LossUnsup$ is shown in \eqref{eq:unsupervised_loss}. A natural question is why the solution of the above learning problem does not work properly for the BSCM problem? 
We endeavor to provide an argument as follows: Let the total number of scalars in $\btheta$ {be} $N_{\theta}$. Assuming $T^{(i)} = T, \forall i \in \IndexN{N}$, the total number of constraints in unsupervised learning is $nNT$. Typically $N_{\theta} < nNT$, and even $N_{\theta} \ll nNT$, hence the learning of $\btheta$ may not be necessarily under-determined. However, learning an appropriate $\btheta$ for state estimation remains an under-determined problem. {We can learn $\btheta$ by optimizing \eqref{eq:DANSE_learning},} and subsequently, RNN provides a set of parameters for the Gaussian prior that is consistent with the measurement/observation system. However, {this} does not {imply the consistency of the learned parameters with the underlying true states.} 
Overall, this phenomenon can be explained conceptually as a limitation of unsupervised learning because we do not have access to any labelled data to provide a suitable regularization in \eqref{eq:DANSE_learning}. 

{{For semi-supervised learning, we have a limited amount of labelled data available as shown in \eqref{eq:semisupervised_data}}. {This leads to the optimization problem in \eqref{eq:SemiDANSE_learning} which can be achieved from \eqref{eq:DANSE_learning} with the additional regularization term $\sum_{j \in \IndexSet_s} \LossSup\left(\bstate^{(j)}_{1:T^{(j)}}, \bmeas^{(j)}_{1:T^{(j)}}; {\btheta}\right)$.}
We believe that SemiDANSE could work by exploiting the available time-wise and dimension-wise correlation in the true state trajectories using {the} limited amount of labelled training data in \eqref{eq:semisupervised_data} -- something that is unavailable for a purely unsupervised learning-based method like DANSE. {We hypothesize that this availability of a limited amount of labelled data enables the method to exploit the available time-wise and dimension-wise correlation in the true state trajectories from $\Dataset_{s}$, and find a suitable set of learnable parameters. In the case of DANSE and DMM, the learning algorithm also finds a suitable set of parameters, but the methods are notably unsupervised, relying on only noisy measurement trajectories from $\Dataset_{s}$.} Note that a hybrid method like KalmanNet, knowing the underlying dynamical model, can also work {if} there are correlations in the true state trajectories. {Finally}, it is worthwhile noting that if there arises a scenario where two distinct state trajectories get mapped to \textit{identical} measurement trajectories, then it would be very challenging for any method to address the BSCM problem satisfactorily. {Without a known dynamical model to guide the state estimation algorithm, SemiDANSE would try to improve upon DANSE by exploiting the limited amount of labelled training data to learn a suitable $\btheta^{\star}$ as per \eqref{eq:SemiDANSE_learning}. While we cannot provide any theoretical guarantees (such as an MMSE performance bound) for SemiDANSE owing to the use of neural networks, we note favourable empirical state estimation performances in the case of a handful of under-determined measurement systems, shown in the next section.}  

\section{Experiments and Results}\label{sec:experimentsandresults}
\begin{figure}[t]
    \centering
    \includegraphics[width=0.5\textwidth]{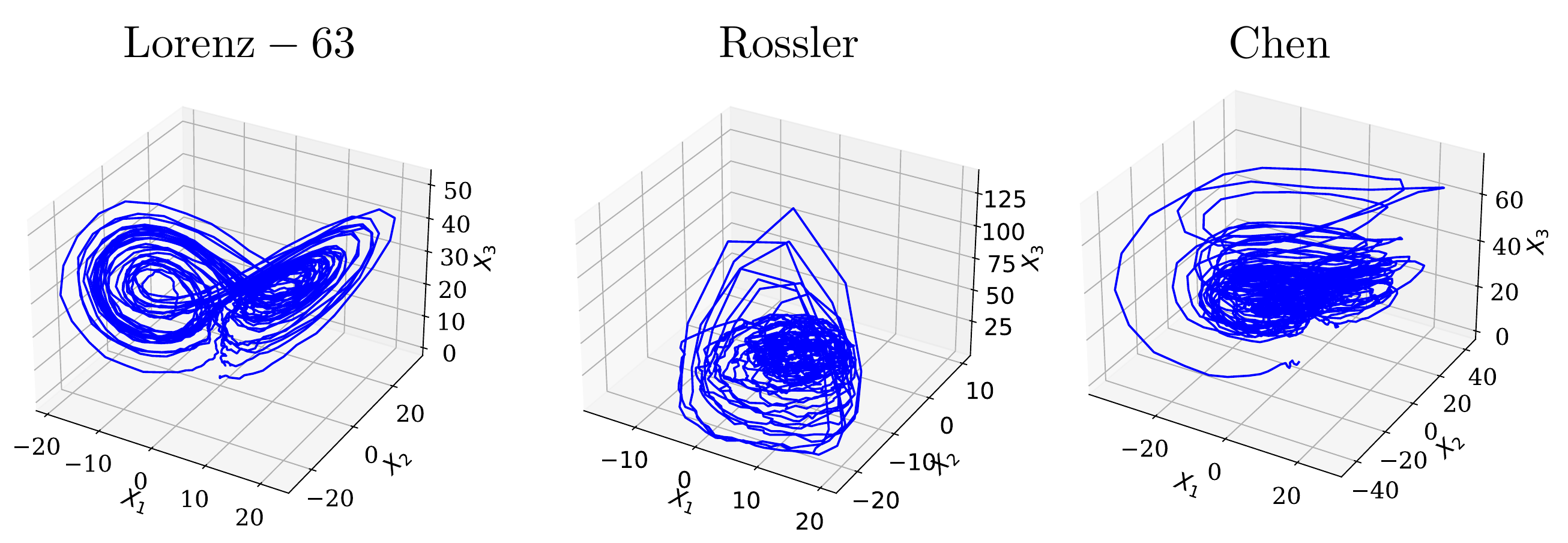}
    \caption{{Visualization of three chaotic dynamical systems: randomly chosen state trajectories from the test set of length $T_{\text{test}}=2000$ for the Lorenz-$63$, Chen, and R\"ossler attractors. The state trajectories show variations of the three systems.}}
    \label{fig:dynamicalsystemsviz}
\end{figure}

{In this section, we demonstrate the empirical performance of SemiDANSE vis-á-vis other Bayesian state estimation methods}. We perform experiments using complex nonlinear processes and generate data knowing their SSMs. While SemiDANSE does not know {the dynamical models} in the SSMs during learning or inference, the {model-driven methods} know the SSMs {(both the dynamical model and the measurement system)}. 

{The organization of this section is as follows: First, we describe three chaotic dynamical systems {(stochastic in nature)} that we simulate as nonlinear processes. Next, we discuss relevant information about the training and testing of SemiDANSE, performance measures and the choice of RNN. We then mention the competing Bayesian state estimation methods and finally illustrate experimental results related to the BSCM problem for the proposed SemiDANSE method.}

\subsection{Three chaotic dynamical systems }\label{sec:chaoticdynsystems}
In this subsection, we briefly describe three chaotic dynamical systems {(stochastic in nature)} used for experiments as benchmark nonlinear systems. They are 
\begin{enumerate}
    \item The Lorenz-$63$ {system} \cite{lorenz1963deterministic}, 
    \item The Chen {system} \cite{chen1999yet}, and 
    \item The R\"ossler {system} \cite{rossler1976equation}.
\end{enumerate}
{In chaos theory literature, the above dynamical systems are often referred to as \textit{strange attractors} \cite{sprott2003chaos}. Hence, we sometimes use the nomenclature `the Lorenz-$63$ attractor' and `the Lorenz-$63$ system' interchangeably.}  These three systems are $3$-dimensional with highly complex state dynamics. We simulate each of the chaotic dynamical systems by discretizing their original continuous-time dynamics following \cite{ghosh2023dansejrnl}. {We also} incorporate i.i.d. process noise to ensure randomness {in the simulated state trajectories}. The detailed equations for their state dynamics {(or dynamical models)} and simulation procedures are described in Appendix \ref{sec:chaoticdynsystems_detail}. All of them are Markovian processes. 
{In all cases, the corresponding noisy measurement trajectories are generated using the linear measurement setup in \eqref{eq:measurement}}.
For visualization, we show randomly chosen test set trajectories of the three processes in Fig. \ref{fig:dynamicalsystemsviz}. This figure helps to visualize how different the trajectories can be across the three chaotic dynamical systems and a qualitative justification for {the use} of them in state estimation experiments across different dynamical systems. {Towards the end of the manuscript, we also experiment using a high-dimensional chaotic dynamical system -- the Lorenz-$96$ system \cite{lorenz1996predictability, lorenz1998optimal}, to explore the applicability of the proposed method.}

\subsection{Training, testing, performance measures and RNN}
\label{sec:trainingtestinghyperparams}
We have training datasets as described in Section \ref{sec:Semi-supervised learning} for DANSE, DMM, and SemiDANSE, and a testing dataset $\Dataset_{\text{test}} = \left\lbrace \left(\bstate^{(j)}_{1:T_{\text{test}}^{(j)}},\bmeas^{(j)}_{1:T_{\text{test}}^{(j)}}\right)\right\rbrace_{j=1}^{\!N_{\text{test}}}$. We mention that SemiDANSE is trained using $\Dataset_{\text{semi}}$, thus involving a combination of labelled and unlabelled data. For training, we use a dataset where $N=1000$ (total number of training trajectories) and having the same length $T=100$ for each training trajectory. For testing, we use a dataset $\Dataset_{\text{test}}$ where $N_{\text{test}}=100$ (total number of testing trajectories) and having the same length $T_{\text{test}}=2000$ for each testing trajectory.
Denoting the estimated state as $\hat{\bstate}_t$ (for example {using the posterior mean as a point estimate}), we use the averaged normalized-mean-squared-error (NMSE) in decibels (dB) as the performance measure, defined below as in \cite[Section III]{ghosh2023dansejrnl}: 
\begin{equation}
    \mathrm{NMSE} = \dfrac{1}{N_{\text{test}}} \sum_{j=1}^{N_{\text{test}}} 10\log_{10}\dfrac{\sum_{t=1}^{T_{\text{test}}^{(j)}}\| \bstate_t^{(j)} - \hat{\bstate}_t^{(j)} \|_2^2}{\sum_{t=1}^{T_{\text{test}}^{(j)}} \| \bstate_t^{(j)}\|_2^2}.
\end{equation}

{For generating noisy measurements,} we use i.i.d., {white} Gaussian noise as the measurement noise {with} $\bmnoiseCov = \sigma_w^2 \mathbf{I}_{\measdim}$. {The process noise is also i.i.d., {white}  Gaussian, with the process noise covariance $\bpnoiseCov$ in most cases being scaled diagonal,  i.e. $\bpnoiseCov = \sigma_e^2 \mathbf{I}_{\statedim}$. Specific details for each kind of {dynamical model} (Lorenz, Chen and Rössler) are given in Appendix \ref{sec:chaoticdynsystems_detail}.} The signal-to-measurement noise ratio (SMNR) is calculated in dB, same as in \cite[Section III]{ghosh2023dansejrnl} on $\Dataset_{\text{test}}$, as
\begin{eqnarray}\label{eq:SMNR}
\begin{array}{rl}
    &\mathrm{SMNR} \!\!\!\!\!\! \\
    &= \dfrac{\sum\limits_{j=1}^{N_{\text{test}}}  \!\! 10 \log_{10} \!\!\left( \sum_{t=1}^{T_{\text{test}}^{(j)}}\dfrac{ \mathbb{E}\lbrace  \| \bmeasmat\bstate_t^{(j)} - \mathbb{E}\lbrace \bmeasmat \bstate_t^{(j)} \rbrace \|_2^2 \rbrace}{\text{tr}\left(\bmnoiseCov\right)}\right)}{N_{\text{test}}} \\
    \!\!&= \dfrac{\sum\limits_{j=1}^{N_{\text{test}}} \!\! 10 \log_{10} \!\! \left( \sum_{t=1}^{T_{\text{test}}^{(j)}}\dfrac{\mathbb{E}\lbrace  \| \bmeasmat\bstate_t^{(j)} - \mathbb{E}\lbrace \bmeasmat \bstate_t^{(j)} \rbrace \|_2^2 \rbrace}{n\sigma_w^2}\right)}{N_{\text{test}}} .
\end{array}
\end{eqnarray}
For notational convenience, we also introduce a parameter 
\begin{eqnarray}\label{eq:kappa}
\fraclabelled \triangleq \dfrac{N_s}{N}, \,\, 0 \leq \fraclabelled \leq 1.    
\end{eqnarray}
The parameter $\fraclabelled$ provides a measure to quantify the relative amount of labelled data and unlabelled data in semi-supervised learning. {As a concrete example, we use $N=1000$ and $N_s = 20$. That means $\fraclabelled = 0.02=2\%$, and SemiDANSE uses $2\%$ labelled data {compared to} the total amount of {labelled and} unlabelled data. Unless the value of $\kappa$ is mentioned specifically, we stick with  $\fraclabelled = 0.02=2\%$ for most of our experiments.} Further, the semi-supervised learning in Section \ref{sec:Semi-supervised learning} is formulated in a way that when $\fraclabelled = 0$ then the semi-supervised learning translates to unsupervised learning. That means, SemiDANSE translates to DANSE when $\fraclabelled = 0$. 

{We found the RNN architecture that includes appropriate feed-forward networks by cross-validation. For SemiDANSE, we experimented with both LSTM and GRU and found the suitable RNN architecture to be a GRU \cite{choPropertiesNeuralMachine2014, hochreiter1997long}. The GRU has $1$ hidden layer with $30$ hidden nodes. The output of the RNN is passed to shallow feed-forward networks $\text{FC}_{\stateMean}, \text{FC}_{\stateCov}$ as described in Fig. \ref{fig:rnnGaussianprior} for calculating the means and the covariances. The feed-forward networks have $2$ hidden layers, where the first hidden layer has $30$ hidden nodes and is shared between $\text{FC}_{\stateMean}$ and $\text{FC}_{\stateCov}$. The output layer was designed to have $32$ hidden nodes and distinct for $\text{FC}_{\stateMean} \text{ and } \text{FC}_{\stateCov}$. The mathematical details of the feed-forward networks are described in \eqref{eq:meancovpriorfeedforward} in Appendix \ref{sec:parameterizationGaussianPriorRNN}}. We implemented SemiDANSE in Python and PyTorch \cite{paszke2019pytorch} and trained the architecture using a single NVIDIA Tesla P100 GPU card \footnote{The code will be made available upon request.}
The training algorithm uses a mini-batch gradient descent with a batch size of $64$. The optimizer chosen was Adam \cite{kingma2014adam} with an adaptive learning rate set at a starting value $5 \times 10^{-4}$ and decreased by $10 \%$ every ${1/6}$'th of the maximum number of training epochs. The maximum number of training epochs was set at $2000$, and an early stopping criterion based on the MSE of the validation set was used to avoid overfitting, similar to DANSE.

\subsection{Competing methods}
In this subsection, we compare SemiDANSE vis-\`a-vis a few other Bayesian state estimation methods like \cite{ghosh2023dansejrnl} on the BSCM task. The methods that we compare are:
\begin{enumerate}
    \item {model-driven} EKF and UKF, 
    \item data-driven causal DMM and DANSE,
    \item hybrid method KalmanNet.
\end{enumerate}
{{All} the above methods (including SemiDANSE) have knowledge of the linear measurement system given by \eqref{eq:measurement}.}
{Additionally, the} {model-driven} EKF and UKF have full knowledge of the {dynamical model of the} underlying process {and are used for benchmarking the performance of data-driven methods}. They are implemented using Python, PyTorch and FilterPy \cite{labbe2020filterpy}. The DMM and DANSE methods are implemented in PyTorch and trained by unsupervised learning that uses $\Dataset_{u}$. DMM assumes that the underlying process is Markovian, but it does not know the {exact dynamical model}. A brief description of the learning problem for the DMM that we implemented in this article is given in Appendix \ref{sec:dmmdetail}. DANSE does not have any knowledge of the underlying process. The KalmanNet is trained by unsupervised learning that uses $\Dataset_{u}$ and it also has the full knowledge of the {process dynamics} \cite{revach2022unsupervised}. {The training and simulation strategies for these competing methods follow the same as described in \cite[Section III]{ghosh2023dansejrnl}.}


\begin{figure}[t]
    \centering
    \scalebox{1.0}{
\begin{tikzpicture}

\definecolor{darkgray176}{RGB}{176,176,176}
\definecolor{darkturquoise0191191}{RGB}{0,191,191}
\definecolor{darkviolet1910191}{RGB}{191,0,191}
\definecolor{goldenrod1911910}{RGB}{191,191,0}
\definecolor{green01270}{RGB}{0,127,0}
\definecolor{lightgray204}{RGB}{204,204,204}

\begin{axis}[
legend cell align={left},
legend style={
  fill opacity=0.8,
  draw opacity=1,
  text opacity=1,
  at={(0.18,0.42)},
  anchor=north,
  draw=lightgray204
},
tick align=outside,
tick pos=left,
x grid style={darkgray176},
xlabel={SMNR (in dB)},
xmajorgrids,
xmin=-12, xmax=32,
xtick style={color=black},
y grid style={darkgray176},
ylabel={NMSE (in dB)},
ymajorgrids,
ymin=-32.9838744260371, ymax=1.1297762401402,
ytick style={color=black},
height=0.35\textwidth,
width=0.49\textwidth
]

\path [draw=red, semithick]
(axis cs:-10,-2.31937265396118)
--(axis cs:-10,-1.43408346176147);

\path [draw=red, semithick]
(axis cs:0,-5.0203475356102)
--(axis cs:0,-3.33978277444839);

\path [draw=red, semithick]
(axis cs:10,-16.7178335189819)
--(axis cs:10,-9.24999523162842);

\path [draw=red, semithick]
(axis cs:20,-26.5824261903763)
--(axis cs:20,-25.7503756284714);

\path [draw=red, semithick]
(axis cs:30,-30.8423448503017)
--(axis cs:30,-30.3979734480381);

\path [draw=black, semithick]
(axis cs:-10,-5.56235510110855)
--(axis cs:-10,-5.00874322652817);

\path [draw=black, semithick]
(axis cs:0,-9.19363224506378)
--(axis cs:0,-6.85526072978973);

\path [draw=black, semithick]
(axis cs:10,-19.2293984889984)
--(axis cs:10,-12.0651075839996);

\path [draw=black, semithick]
(axis cs:20,-26.4918423891068)
--(axis cs:20,-25.6642359495163);

\path [draw=black, semithick]
(axis cs:30,-30.7380625605583)
--(axis cs:30,-30.3109791874886);
\path [draw=darkturquoise0191191, semithick]
(axis cs:-10,-3.84509596973658)
--(axis cs:-10,-3.64178528636694);

\path [draw=darkturquoise0191191, semithick]
(axis cs:0,-2.80664260685444)
--(axis cs:0,-2.5594195574522);

\path [draw=darkturquoise0191191, semithick]
(axis cs:10,-2.29420465976)
--(axis cs:10,-2.05728154629469);

\path [draw=darkturquoise0191191, semithick]
(axis cs:20,-1.93166395276785)
--(axis cs:20,-1.70128969103098);

\path [draw=darkturquoise0191191, semithick]
(axis cs:30,-1.81230589747429)
--(axis cs:30,-1.65008881688118);

\path [draw=blue, semithick]
(axis cs:-10,-1.32919542491436)
--(axis cs:-10,-1.01925428211689);

\path [draw=blue, semithick]
(axis cs:0,-3.43626280128956)
--(axis cs:0,-3.05725984275341);

\path [draw=blue, semithick]
(axis cs:10,-3.39542695879936)
--(axis cs:10,-2.96321561932564);

\path [draw=blue, semithick]
(axis cs:20,-3.83723993599415)
--(axis cs:20,-3.41399984061718);

\path [draw=blue, semithick]
(axis cs:30,-4.50034965574741)
--(axis cs:30,-4.07380805909634);
\addplot [semithick, red, dashed, mark=diamond*, mark size=3, mark options={solid}]
table {%
-10 -1.87672805786133
0 -4.1800651550293
10 -12.9839143753052
20 -26.1664009094238
30 -30.6201591491699
};
\addlegendentry{EKF}
\addplot [semithick, black, dashed, mark=*, mark size=3, mark options={solid}]
table {%
-10 -5.28554916381836
0 -8.02444648742676
10 -15.647253036499
20 -26.0780391693115
30 -30.5245208740234
};
\addlegendentry{UKF}
\addplot [semithick, darkturquoise0191191, mark=square, mark size=3, mark options={solid}]
table {%
-10 -3.74344062805176
0 -2.68303108215332
10 -2.17574310302734
20 -1.81647682189941
30 -1.73119735717773
};
\addlegendentry{DMM}
\addplot [semithick, blue, mark=triangle*, mark size=3, mark options={solid}]
table {%
-10 -1.17422485351562
0 -3.24676132202148
10 -3.1793212890625
20 -3.62561988830566
30 -4.28707885742188
};
\addlegendentry{DANSE}
\end{axis}

\end{tikzpicture}}
    \caption{The average NMSE (in dB) on $\Dataset_{\text{test}}$ versus SMNR (in dB) performances to illustrate failure of unsupervised learning-based DANSE and DMM for the BSCM problem setup described in Section \ref{subsec:failurecase}, {with $\sigma_{\pnoise}^2$ corresponding to $-10$ dB}. The model-driven EKF and UKF perform well. 
    }
\label{fig:nmse_vs_smnr_unstructured_measurements_failurecase}
\end{figure}

\subsection{Failure of unsupervised DANSE and DMM for BSCM}
\label{subsec:failurecase}
We begin our experiments with {the} Lorenz-$63$ system, where we first show the failure of {two} unsupervised learning-based methods - DANSE and DMM {for the BSCM task}. For the experiment, we use $2 \times 3$-dimensional $\bmeasmat$ matrix where each element is drawn from i.i.d. Gaussian source $\mathcal{N}(0,1)$ and then fixed. Therefore we have $n=2$ and $m=3$, and the $\bmeasmat$ matrix we use for the BSCM problem is
\begin{eqnarray}
    \bmeasmat= \begin{bmatrix}
        0.37992 &  0.34099 &  1.04317 \\
        0.98070 & -0.70477 &  2.17908 \\
    \end{bmatrix}.
\end{eqnarray}
In Fig. \ref{fig:nmse_vs_smnr_unstructured_measurements_failurecase}, we show the NMSE versus SMNR performances of DANSE and DMM, and only include the performances of EKF and UKF for comparison. We note that both DANSE and DMM were trained on different SMNRs as shown in Fig. \ref{fig:nmse_vs_smnr_unstructured_measurements_failurecase} and tested on the same SMNRs. It shows that the model-driven EKF and UKF perform well, but the unsupervised learning-based DANSE and DMM - both of them - fail to perform {satisfactorily}. Note that the Lorenz-$63$ system that we simulated is a Markovian process and DMM uses the knowledge that the underlying dynamical model is Markovian. {Despite that, DMM is unable} to address the BSCM problem.

Here, for completeness, we mention that if the $\mathbf{H}$ matrix would be a full column-rank matrix, for example, a $3 \times 3$-dimensional full-rank matrix, then DANSE and DMM could perform well, {see} \cite[Section III-D]{ghosh2023dansejrnl}. 
\begin{figure}[t]
    \centering
    \scalebox{1.0}{
\begin{tikzpicture}

\definecolor{darkgray176}{RGB}{176,176,176}
\definecolor{darkturquoise0191191}{RGB}{0,191,191}
\definecolor{darkviolet1910191}{RGB}{191,0,191}
\definecolor{goldenrod1911910}{RGB}{191,191,0}
\definecolor{green01270}{RGB}{0,127,0}
\definecolor{lightgray204}{RGB}{204,204,204}

\begin{axis}[
legend cell align={left},
legend style={
  fill opacity=0.8,
  draw opacity=1,
  text opacity=1,
  at={(0.24,0.62)},
  anchor=north,
  draw=lightgray204
},
tick align=outside,
tick pos=left,
x grid style={darkgray176},
xlabel={SMNR (in dB)},
xmajorgrids,
xmin=-12, xmax=32,
xtick style={color=black},
y grid style={darkgray176},
ylabel={NMSE (in dB)},
ymajorgrids,
ymin=-32.9838744260371, ymax=1.1297762401402,
ytick style={color=black},
height=0.38\textwidth,
width=0.49\textwidth
]
\path [draw=red, semithick]
(axis cs:-10,-2.31937265396118)
--(axis cs:-10,-1.43408346176147);

\path [draw=red, semithick]
(axis cs:0,-5.0203475356102)
--(axis cs:0,-3.33978277444839);

\path [draw=red, semithick]
(axis cs:10,-16.7178335189819)
--(axis cs:10,-9.24999523162842);

\path [draw=red, semithick]
(axis cs:20,-26.5824261903763)
--(axis cs:20,-25.7503756284714);

\path [draw=red, semithick]
(axis cs:30,-30.8423448503017)
--(axis cs:30,-30.3979734480381);

\path [draw=black, semithick]
(axis cs:-10,-5.56235510110855)
--(axis cs:-10,-5.00874322652817);

\path [draw=black, semithick]
(axis cs:0,-9.19363224506378)
--(axis cs:0,-6.85526072978973);

\path [draw=black, semithick]
(axis cs:10,-19.2293984889984)
--(axis cs:10,-12.0651075839996);

\path [draw=black, semithick]
(axis cs:20,-26.4918423891068)
--(axis cs:20,-25.6642359495163);

\path [draw=black, semithick]
(axis cs:30,-30.7380625605583)
--(axis cs:30,-30.3109791874886);

\path [draw=goldenrod1911910, semithick]
(axis cs:-10,-2.92340674996376)
--(axis cs:-10,-2.29357704520226);

\path [draw=goldenrod1911910, semithick]
(axis cs:0,-4.55577248334885)
--(axis cs:0,-3.38844138383865);

\path [draw=goldenrod1911910, semithick]
(axis cs:10,-9.07130831480026)
--(axis cs:10,-7.12318783998489);

\path [draw=goldenrod1911910, semithick]
(axis cs:20,-20.9247599840164)
--(axis cs:20,-19.0170048475266);

\path [draw=goldenrod1911910, semithick]
(axis cs:30,-26.442945882678)
--(axis cs:30,-25.9855666905642);

\path [draw=blue, semithick]
(axis cs:-10,-1.32919542491436)
--(axis cs:-10,-1.01925428211689);

\path [draw=blue, semithick]
(axis cs:0,-3.43626280128956)
--(axis cs:0,-3.05725984275341);

\path [draw=blue, semithick]
(axis cs:10,-3.39542695879936)
--(axis cs:10,-2.96321561932564);

\path [draw=blue, semithick]
(axis cs:20,-3.83723993599415)
--(axis cs:20,-3.41399984061718);

\path [draw=blue, semithick]
(axis cs:30,-4.50034965574741)
--(axis cs:30,-4.07380805909634);

\path [draw=darkviolet1910191, semithick]
(axis cs:-10,-5.69188467413187)
--(axis cs:-10,-5.45723184198141);

\path [draw=darkviolet1910191, semithick]
(axis cs:0,-6.62664254754782)
--(axis cs:0,-6.45554510504007);

\path [draw=darkviolet1910191, semithick]
(axis cs:10,-14.0616897344589)
--(axis cs:10,-13.2925812005997);

\path [draw=darkviolet1910191, semithick]
(axis cs:20,-22.7293455898762)
--(axis cs:20,-21.7876450717449);

\path [draw=darkviolet1910191, semithick]
(axis cs:30,-29.0430081486702)
--(axis cs:30,-28.3118922114372);
\addplot [semithick, red, dashed, mark=diamond*, mark size=3, mark options={solid}]
table {%
-10 -1.87672805786133
0 -4.1800651550293
10 -12.9839143753052
20 -26.1664009094238
30 -30.6201591491699
};
\addlegendentry{EKF}
\addplot [semithick, black, dashed, mark=*, mark size=3, mark options={solid}]
table {%
-10 -5.28554916381836
0 -8.02444648742676
10 -15.647253036499
20 -26.0780391693115
30 -30.5245208740234
};
\addlegendentry{UKF}
\addplot [semithick, goldenrod1911910, mark=square*, mark size=3, mark options={solid}]
table {%
-10 -2.60849189758301
0 -3.97210693359375
10 -8.09724807739258
20 -19.9708824157715
30 -26.2142562866211
};
\addlegendentry{KalmanNet}
\addplot [semithick, darkturquoise0191191, mark=square, mark size=3, mark options={solid}]
table {%
-10 -3.74344062805176
0 -2.68303108215332
10 -2.17574310302734
20 -1.81647682189941
30 -1.73119735717773
};
\addlegendentry{{DMM}}
\addplot [semithick, blue, mark=triangle*, mark size=3, mark options={solid}]
table {%
-10 -1.17422485351562
0 -3.24676132202148
10 -3.1793212890625
20 -3.62561988830566
30 -4.28707885742188
};
\addlegendentry{DANSE}
\addplot [semithick, darkviolet1910191, mark=pentagon, mark size=3, mark options={solid}]
table {%
-10 -5.57455825805664
0 -6.54109382629395
10 -13.6771354675293
20 -22.2584953308105
30 -28.6774501800537
};
\addlegendentry{SemiDANSE}
\end{axis}

\end{tikzpicture}}
    \caption{{The average NMSE (in dB) on $\Dataset_{\text{test}}$ versus SMNR (in dB) performances to illustrate the success of SemiDANSE for the BSCM problem setup described in Section \ref{subsec:failurecase}, {with $\sigma_{\pnoise}^2$ corresponding to $-10$ dB}. SemiDANSE ($\fraclabelled=0.02$) is compared with the model-driven EKF and UKF, the DMM \cite{krishnan2017structured}, the hybrid KalmanNet \cite{revach2022unsupervised} and the unsupervised learning-based DANSE.}}
    \label{fig:nmse_vs_smnr_unstructured_measurements}
\end{figure}

\subsection{Success of SemiDANSE for BSCM}\label{sec:semidanse_success}
For the same setup of the Section \ref{subsec:failurecase}, we now show the performance of SemiDANSE and compare it with EKF, UKF and KalmanNet in Fig. \ref{fig:nmse_vs_smnr_unstructured_measurements}. As in Section \ref{subsec:failurecase}, SemiDANSE was also trained and tested on different SMNRs using $\fraclabelled=0.02$ to keep the comparison consistent. We observe that SemiDANSE is competitive {and} overcomes the limitation of DANSE for the BSCM problem. {We remark that} using $2\%$ labelled data {brings forth the desired regularization in SemiDANSE, leading to an improved state estimation performance}. {Additionally, we empirically provide an ablation study for assesing the individual contributions of the semi-supervised loss, namely the supervised and the unsupervised loss terms.
\begin{figure}[t]
    \centering
    \scalebox{1.0}{
\begin{tikzpicture}

\definecolor{darkgray176}{RGB}{176,176,176}
\definecolor{darkturquoise0191191}{RGB}{0,191,191}
\definecolor{darkviolet1910191}{RGB}{191,0,191}
\definecolor{goldenrod1911910}{RGB}{191,191,0}
\definecolor{green01270}{RGB}{0,127,0}
\definecolor{lightgray204}{RGB}{204,204,204}

\begin{axis}[
legend cell align={left},
legend style={
  fill opacity=0.8,
  draw opacity=1,
  text opacity=1,
  at={(0.3,0.41)},
  anchor=north,
  draw=lightgray204
},
tick align=outside,
tick pos=left,
x grid style={darkgray176},
xlabel={SMNR (in dB)},
xmajorgrids,
xmin=-12, xmax=32,
xtick style={color=black},
y grid style={darkgray176},
ylabel={NMSE (in dB)},
ymajorgrids,
ymin=-32.9759920403361, ymax=1.114662335813,
ytick style={color=black},
width=\linewidth,
height=0.6\linewidth
]
\path [draw=blue, semithick]
(axis cs:-10,-1.32889050245285)
--(axis cs:-10,-1.03654223680496);

\path [draw=blue, semithick]
(axis cs:0,-3.48227868974209)
--(axis cs:0,-2.99839986860752);

\path [draw=blue, semithick]
(axis cs:10,-3.38209737837315)
--(axis cs:10,-2.97856698930264);

\path [draw=blue, semithick]
(axis cs:20,-3.82639230787754)
--(axis cs:20,-3.4243019670248);

\path [draw=blue, semithick]
(axis cs:30,-4.46318323910236)
--(axis cs:30,-4.05587880313396);

\path [draw=darkviolet1910191, semithick]
(axis cs:-10,-3.84866113960743)
--(axis cs:-10,-3.63705663383007);

\path [draw=darkviolet1910191, semithick]
(axis cs:0,-2.81636948883533)
--(axis cs:0,-2.53963713347912);

\path [draw=darkviolet1910191, semithick]
(axis cs:10,-2.29096639156342)
--(axis cs:10,-2.06006968021393);

\path [draw=darkviolet1910191, semithick]
(axis cs:20,-1.92886895686388)
--(axis cs:20,-1.70995550602674);

\path [draw=darkviolet1910191, semithick]
(axis cs:30,-1.79979556798935)
--(axis cs:30,-1.64635235071182);

\path [draw=darkturquoise0191191, semithick]
(axis cs:-10,-5.73141155391932)
--(axis cs:-10,-5.49824085086584);

\path [draw=darkturquoise0191191, semithick]
(axis cs:0,-3.58799386024475)
--(axis cs:0,-2.64163565635681);

\path [draw=darkturquoise0191191, semithick]
(axis cs:10,-8.92109894752502)
--(axis cs:10,-5.92457747459412);

\path [draw=darkturquoise0191191, semithick]
(axis cs:20,-18.8898546695709)
--(axis cs:20,-17.4696080684662);

\path [draw=darkturquoise0191191, semithick]
(axis cs:30,-25.4874188899994)
--(axis cs:30,-23.7615168094635);

\path [draw=darkviolet1910191, semithick]
(axis cs:-10,-5.69988595694304)
--(axis cs:-10,-5.47766912728548);

\path [draw=darkviolet1910191, semithick]
(axis cs:0,-6.64788433909416)
--(axis cs:0,-6.44421008229256);

\path [draw=darkviolet1910191, semithick]
(axis cs:10,-14.015333712101)
--(axis cs:10,-13.2903684973717);

\path [draw=darkviolet1910191, semithick]
(axis cs:20,-22.7768052816391)
--(axis cs:20,-21.7693411111832);

\path [draw=darkviolet1910191, semithick]
(axis cs:30,-29.0517361760139)
--(axis cs:30,-28.1476199030876);
\addplot [semithick, blue, mark=triangle*, mark size=3, mark options={solid}]
table {%
-10 -1.18271636962891
0 -3.2403392791748
10 -3.18033218383789
20 -3.62534713745117
30 -4.25953102111816
};
\addlegendentry{DANSE}
\addplot [semithick, darkturquoise0191191, mark=pentagon, mark size=3, mark options={solid}]
table {%
-10 -5.61482620239258
0 -3.11481475830078
10 -7.42283821105957
20 -18.1797313690186
30 -24.6244678497314
};
\addlegendentry{Supervised loss}
\addplot [semithick, darkviolet1910191, mark=pentagon, mark size=3, mark options={solid}]
table {%
-10 -5.58877754211426
0 -6.54604721069336
10 -13.6528511047363
20 -22.2730731964111
30 -28.5996780395508
};
\addlegendentry{SemiDANSE}
\end{axis}

\end{tikzpicture}}
    \caption{{The average NMSE (in dB) on $\Dataset_{\text{test}}$ versus SMNR (in dB) performances to illustrate the success of SemiDANSE ($\fraclabelled=0.02$) for the BSCM problem setup for the BSCM problem setup described in Section \ref{subsec:failurecase}, {with $\sigma_{\pnoise}^2$ corresponding to $-10$ dB}. SemiDANSE ($\fraclabelled=0.02$) is compared with the unsupervised learning-based DANSE and the supervised baseline using only $\LossSup$ in \eqref{eq:supervised_loss}.}}
    \label{fig:nmse_vs_smnr_partial_lorenz63_ablation}
\end{figure}
The results are shown in Fig. \ref{fig:nmse_vs_smnr_partial_lorenz63_ablation}. SemiDANSE is empirically found to perform better than a supervised baseline that is optimized using only $\LossSup$ as shown in \eqref{eq:supervised_loss}.} 

\begin{figure}[t]
    \centering
    \includegraphics[width=0.49\textwidth]{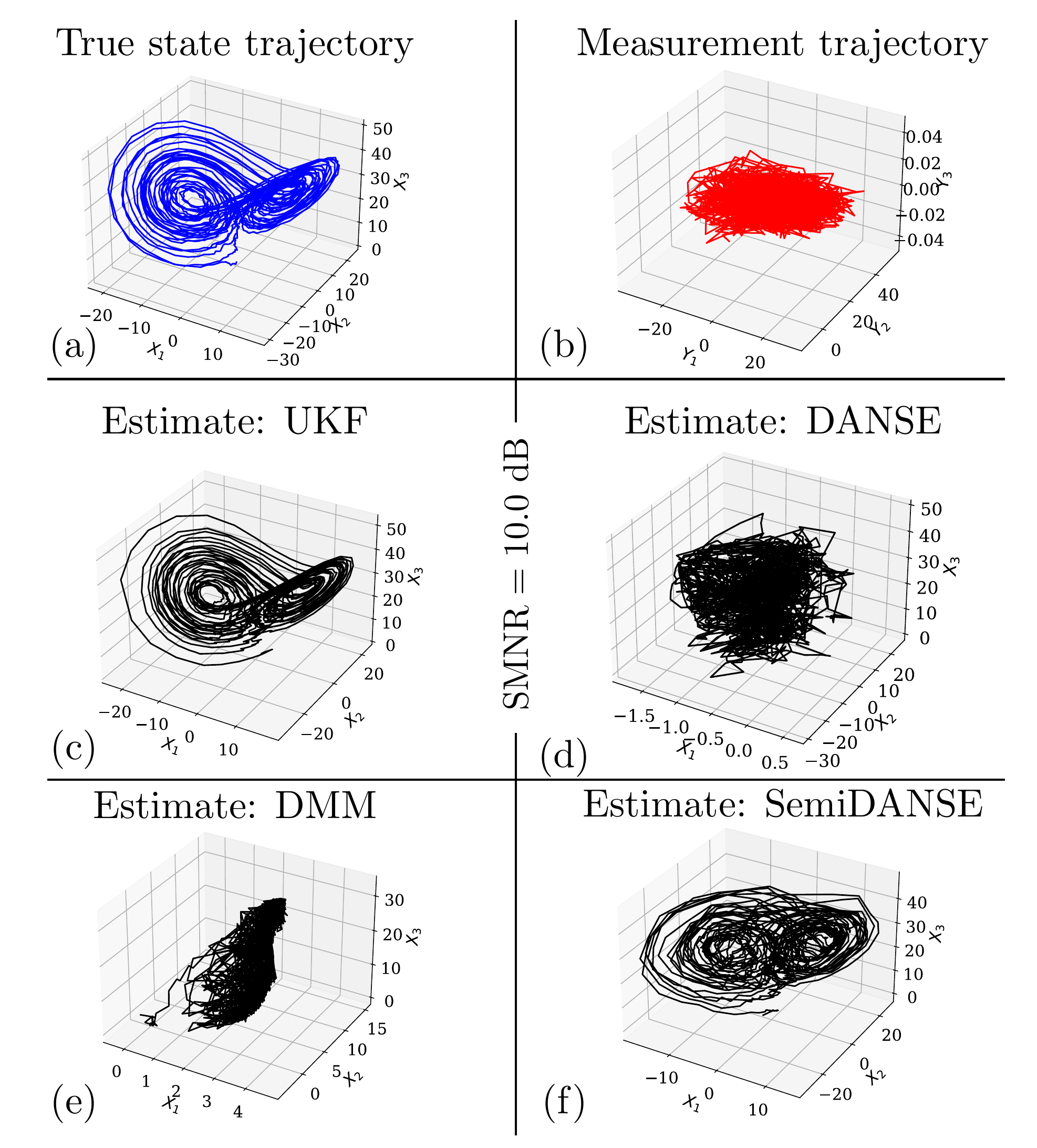}
    \caption{{Demonstrating failure of DANSE and success of SemiDANSE for BSCM problem with {the} partial measurement of Lorenz-$63$ process. We use the BSCM problem setup described in Section \ref{subsec:PartialMeasurement}, where $\mathbf{H}$ is shown in \eqref{eq:Partial_H}, $\text{SMNR} = 10 \text{ dB}$ and $\sigmapnoise^2$ corresponding to $-10$ dB. (a) A $3$-dimensional true state trajectory instance of Lorenz-$63$ system from $\Dataset_{\text{test}}$. (b) The corresponding $2$-dimensional (noisy) measurement trajectory (shown in three-dimensions for visualization). (c) The estimate of UKF. (d) The estimate of DMM. (e) The estimate of DANSE (posterior mean). (f) The estimate of SemiDANSE (posterior mean) with $\fraclabelled=0.02$.}}
    \label{fig:semidanse_demonstrator}
\end{figure}

{\begin{figure}[t]
    \centering
    \scalebox{1.0}{\input{figs/revision_2/Axes_w_lims_sigmae2_-10.0dB_smnr_10.0dB}}
    \caption{{Time-wise plot of a (true) state trajectory $x_{t,1}$ from $\Dataset_{\text{test}}$ and its estimates using UKF, DANSE, DMM and SemiDANSE ($\fraclabelled=0.02$) at $\text{SMNR} = 10$ dB and $\sigmapnoise^2$ corresponding to $-10$ dB for the Lorenz-$63$ process. The results are for the BSCM problem setup described in Section \ref{subsec:PartialMeasurement}. The solid/dashed line represents the corresponding posterior mean estimate. Note that DMM and DANSE fail to track the true state, whereas UKF and SemiDANSE can track it. A short portion of the test trajectory is shown for ease of visualization. 
    }}
    \label{fig:axeswise_plot_LorenzSSMn2}
\end{figure}
}
\subsection{Partial measurement system}
\label{subsec:PartialMeasurement}
In the previous two subsections, we had a dense $\mathbf{H}$ matrix providing the $2$-dimensional measurement vector $\bmeas_t$ that has influences from all the three components of the state vector $\bstate_t$. Instead of a dense $\mathbf{H}$, we can have scenarios where only some components of the state vectors are measured (observed). We refer to these scenarios as partial measurement scenarios.

Let us assume that a partial measurement matrix is
\begin{eqnarray}
\label{eq:Partial_H}
    \bmeasmat= \begin{bmatrix}
        0 &  1 & 0 \\
        0 & 0 &  1 \\
    \end{bmatrix}.
\end{eqnarray}
That means we have measurements from the second and third components of the $3$-dimensional state vector. Fig. \ref{fig:semidanse_demonstrator} shows a $3$-dimensional trajectory of the Lorenz-$63$ process, its $2$-dimensional measurement at SMNR $= 10$ dB, the estimated trajectory using DANSE, and the same using SemiDANSE. It is clear that DANSE fails for the BSCM problem and SemiDANSE performs well. 
\begin{figure}[t]
    \centering
    \scalebox{1.0}{
\begin{tikzpicture}

\definecolor{darkgray176}{RGB}{176,176,176}
\definecolor{darkturquoise0191191}{RGB}{0,191,191}
\definecolor{goldenrod1911910}{RGB}{191,191,0}
\definecolor{lightgray204}{RGB}{204,204,204}

\begin{axis}[
legend cell align={left},
legend style={fill opacity=0.8, draw opacity=1, text opacity=1, draw=lightgray204},
tick align=outside,
tick pos=left,
x grid style={darkgray176},
xlabel={{SMNR (in dB)}},
xmajorgrids,
xmin=-12, xmax=32,
xtick style={color=black},
xtick={-10,0,10,20,30},
xticklabels={
  \(\displaystyle {\ensuremath{-}10}\),
  \(\displaystyle {0}\),
  \(\displaystyle {10}\),
  \(\displaystyle {20}\),
  \(\displaystyle {30}\),
},
y grid style={darkgray176},
ylabel={{NMSE (in dB)}},
ymajorgrids,
ymin=-32.6351144321263, ymax=-3.52564965113997,
ytick style={color=black},
ytick={-35,-30,-25,-20,-15,-10,-5,0},
yticklabels={
  \(\displaystyle {\ensuremath{-}35}\),
  \(\displaystyle {\ensuremath{-}30}\),
  \(\displaystyle {\ensuremath{-}25}\),
  \(\displaystyle {\ensuremath{-}20}\),
  \(\displaystyle {\ensuremath{-}15}\),
  \(\displaystyle {\ensuremath{-}10}\),
  \(\displaystyle {\ensuremath{-}5}\),
  \(\displaystyle {0}\)
},
width=0.49\textwidth,
height=0.3\textwidth
]
\path [draw=blue, semithick]
(axis cs:-10,-5.23542137444019)
--(axis cs:-10,-4.84880714118481);

\path [draw=blue, semithick]
(axis cs:0,-10.3646919727325)
--(axis cs:0,-9.79563212394714);

\path [draw=blue, semithick]
(axis cs:10,-14.7104110717773)
--(axis cs:10,-14.3220462799072);

\path [draw=blue, semithick]
(axis cs:20,-17.1402774006128)
--(axis cs:20,-16.7031292766333);

\path [draw=blue, semithick]
(axis cs:30,-18.8201200664043)
--(axis cs:30,-18.2314355671406);

\path [draw=goldenrod1911910, semithick]
(axis cs:-10,-5.8873470723629)
--(axis cs:-10,-5.61867251992226);

\path [draw=goldenrod1911910, semithick]
(axis cs:0,-8.35058596730232)
--(axis cs:0,-8.0316581428051);

\path [draw=goldenrod1911910, semithick]
(axis cs:10,-16.2811941951513)
--(axis cs:10,-15.8094506412745);

\path [draw=goldenrod1911910, semithick]
(axis cs:20,-23.7289849072695)
--(axis cs:20,-23.2321585863829);

\path [draw=goldenrod1911910, semithick]
(axis cs:30,-30.3457446396351)
--(axis cs:30,-29.7844891250134);

\path [draw=darkturquoise0191191, semithick]
(axis cs:-10,-5.87372357398272)
--(axis cs:-10,-5.62832491844893);

\path [draw=darkturquoise0191191, semithick]
(axis cs:0,-10.6831774115562)
--(axis cs:0,-9.1729970574379);

\path [draw=darkturquoise0191191, semithick]
(axis cs:10,-19.2543579936028)
--(axis cs:10,-18.2475913167);

\path [draw=darkturquoise0191191, semithick]
(axis cs:20,-27.1938781291246)
--(axis cs:20,-26.7566605061293);

\path [draw=darkturquoise0191191, semithick]
(axis cs:30,-31.3119569420815)
--(axis cs:30,-30.827008664608);

\addplot [semithick, line width=1, blue, mark=triangle, mark size=4, mark options={solid,rotate=180}]
table {%
-10 -5.0421142578125
0 -10.0801620483398
10 -14.5162286758423
20 -16.921703338623
30 -18.5257778167725
};
\addlegendentry{{$ \kappa=0.002$}}
\addplot [semithick, line width=1, goldenrod1911910, mark=square, mark size=4, mark options={solid}]
table {%
-10 -5.75300979614258
0 -8.19112205505371
10 -16.0453224182129
20 -23.4805717468262
30 -30.0651168823242
};
\addlegendentry{{$ \kappa=0.02$}}
\addplot [semithick, line width=1, darkturquoise0191191, mark=pentagon, mark size=4, mark options={solid}]
table {%
-10 -5.75102424621582
0 -9.92808723449707
10 -18.7509746551514
20 -26.975269317627
30 -31.0694828033447
};
\addlegendentry{{$ \kappa=0.05$}}
\end{axis}

\end{tikzpicture}}
    \caption{{NMSE (in dB) vs. SMNR (in dB) for SemiDANSE on $\Dataset_{\text{test}}$ for the partial measurement case in \ref{subsec:PartialMeasurement} different values of $\kappa = N_s / N$, {with $\sigma_{\pnoise}^2$ corresponding to $-10$ dB.}}}
    \label{fig:LorenzSSMn2_nmse_vs_smnr_diff_kappa}
\end{figure}

\begin{figure}[t]
    \centering
  \subfloat[Plot of $\meas_{t,1}$ versus $t$]{%
       \scalebox{1.0}{\input{figs/Meas_y_lims_LorenzSSMn2_sigmae2_-10.0dB_smnr_10.0dBdim_1}}
        \label{fig:predictiveperformance_a}
    }
    \hfill
    \\
  \subfloat[Plot of $\meas_{t,2}$ versus $t$]{%
        \scalebox{1.0}{\input{figs/Meas_y_lims_LorenzSSMn2_sigmae2_-10.0dB_smnr_10.0dBdim_2}}
         \label{fig:predictiveperformance_b}
    }
  \caption{Time-wise plots of the coordinates of $\bmeas_{t} $ from $\Dataset_{\text{test}}$ and the one-step ahead predicted estimates using DANSE and SemiDANSE $\left(\fraclabelled=0.02\right)$ at $\text{SMNR} = 10$ dB and $\sigmapnoise^2$ corresponding to $-10$ dB. The results are related to the partial measurement system described in subsections $\ref{subsec:PartialMeasurement}$, $\ref{subsec:predictiveperformance}$ and for the Lorenz-$63$ process. The shaded region in each plot shows the $\pm 1\sigma$ point uncertainty of the corresponding predicted estimated of $\lbrace \bmeas_{t} \rbrace$ .
}
\label{fig:predictiveperformance} 
\end{figure}

In Fig. \ref{fig:axeswise_plot_LorenzSSMn2}, we show the time trajectory of the first component $x_{t,1}$ of $\bstate_t$, and estimates of UKF, DANSE and SemiDANSE. DANSE fails to track $x_{t,1}$, and we observe that the model-driven UKF and the data-driven SemiDANSE perform well. 

Note that the $2$-dimensional measurement $\bmeas_t$ has observations of $x_{t,2}$ and $x_{t,3}$ under noise. From our experiments, we have seen that while DANSE fails to estimate $x_{t,1}$, it is able to estimate $x_{t,2}$ and $x_{t,3}$. We do not show the plots of the estimates of $x_{t,2}$ and $x_{t,3}$ for brevity. On the other hand, SemiDANSE could track all three components of $\bstate_t$. 

{Before proceeding to show further results, we seek to analyse the performance of SemiDANSE in case of the partial measurement scenario for different $\kappa$ (defined in \eqref{eq:kappa}) 
}
{To illustrate the behaviour of SemiDANSE for different values of $\kappa$, we plot the NMSE vs. SMNR for three different values of $\kappa=0.002, 0.02 \text{ and } 0.05$ in Fig. \ref{fig:LorenzSSMn2_nmse_vs_smnr_diff_kappa} for the partial measurement scenario described in section \ref{subsec:PartialMeasurement}. We observe that there is a noticeable gain in the estimation performance moving from $\kappa=0.002$ to $\kappa=0.02$, and even minor improvement from $\kappa=0.02$ to $\kappa=0.05$ in certain SMNR values, indicating the benefit of using labelled data in a semi-supervised setup. }
\begin{figure}[t]
    \centering
    \includegraphics[width=0.49\textwidth]{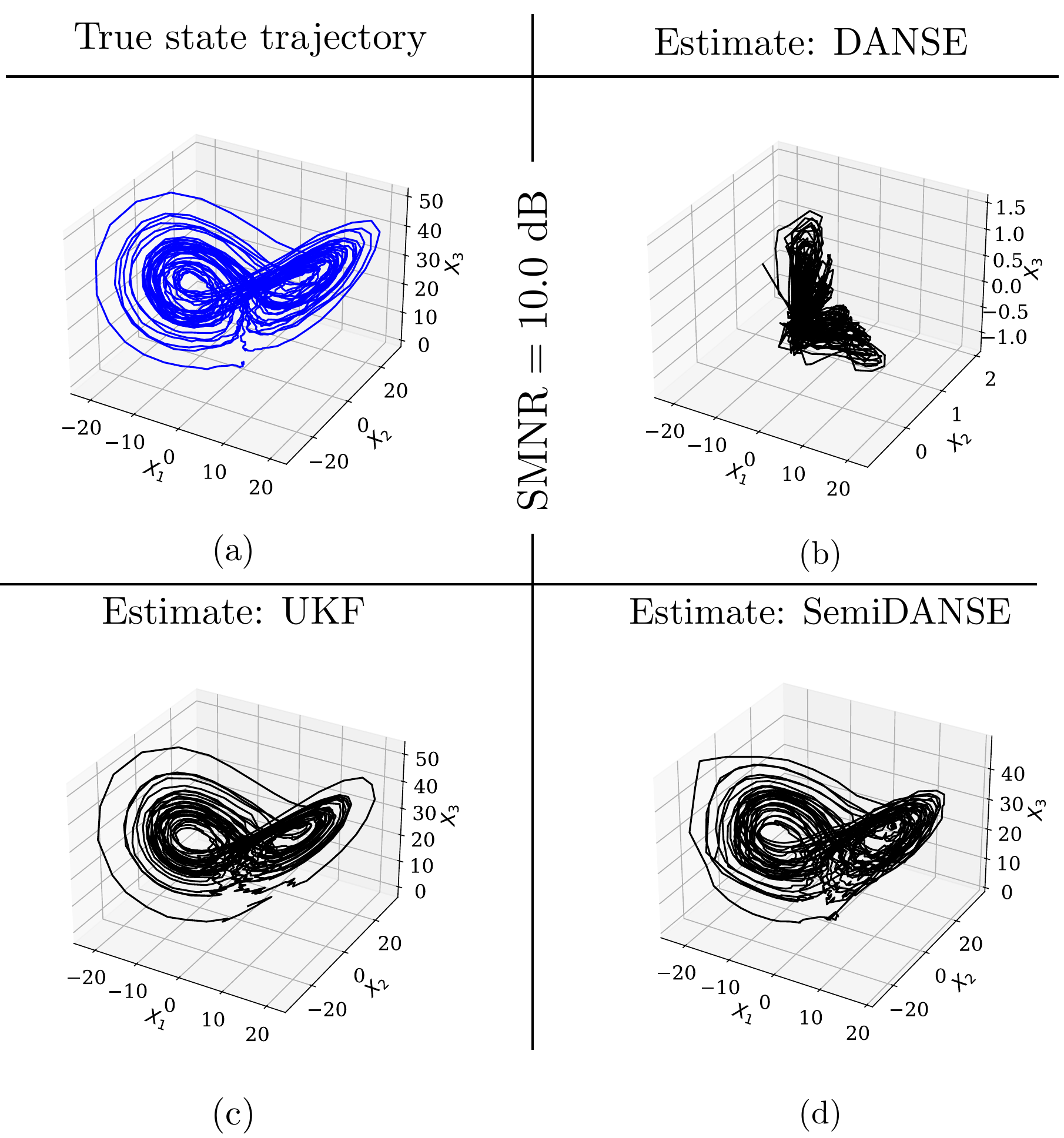}
    \caption{Demonstrating failure of DANSE and success of SemiDANSE for the extreme BSCM problem with one-dimensional partial measurement of $3$-dimensional Lorenz-$63$ process. We use the BSCM problem setup described in Section \ref{subsec:PartialMeasurementExtreme}, where the $1 \times 3$ measurement matrix $\mathbf{H}$ is shown in \eqref{eq:Hn1}, $\text{SMNR} = 10 \text{ dB}$ and $\sigmapnoise^2$ corresponding to $-10$ dB. A $3$-dimensional true state trajectory from $\Dataset_{\text{test}}$ is shown in (a). The estimates for DANSE, UKF and SemiDANSE ($\fraclabelled=0.02$) are all posterior mean estimates shown in (b), (c) and (d).}
    \label{fig:semidanse_demonstrator_n1}
\end{figure}
\begin{figure}[t]
    \centering
    \includegraphics[width=0.5\textwidth]{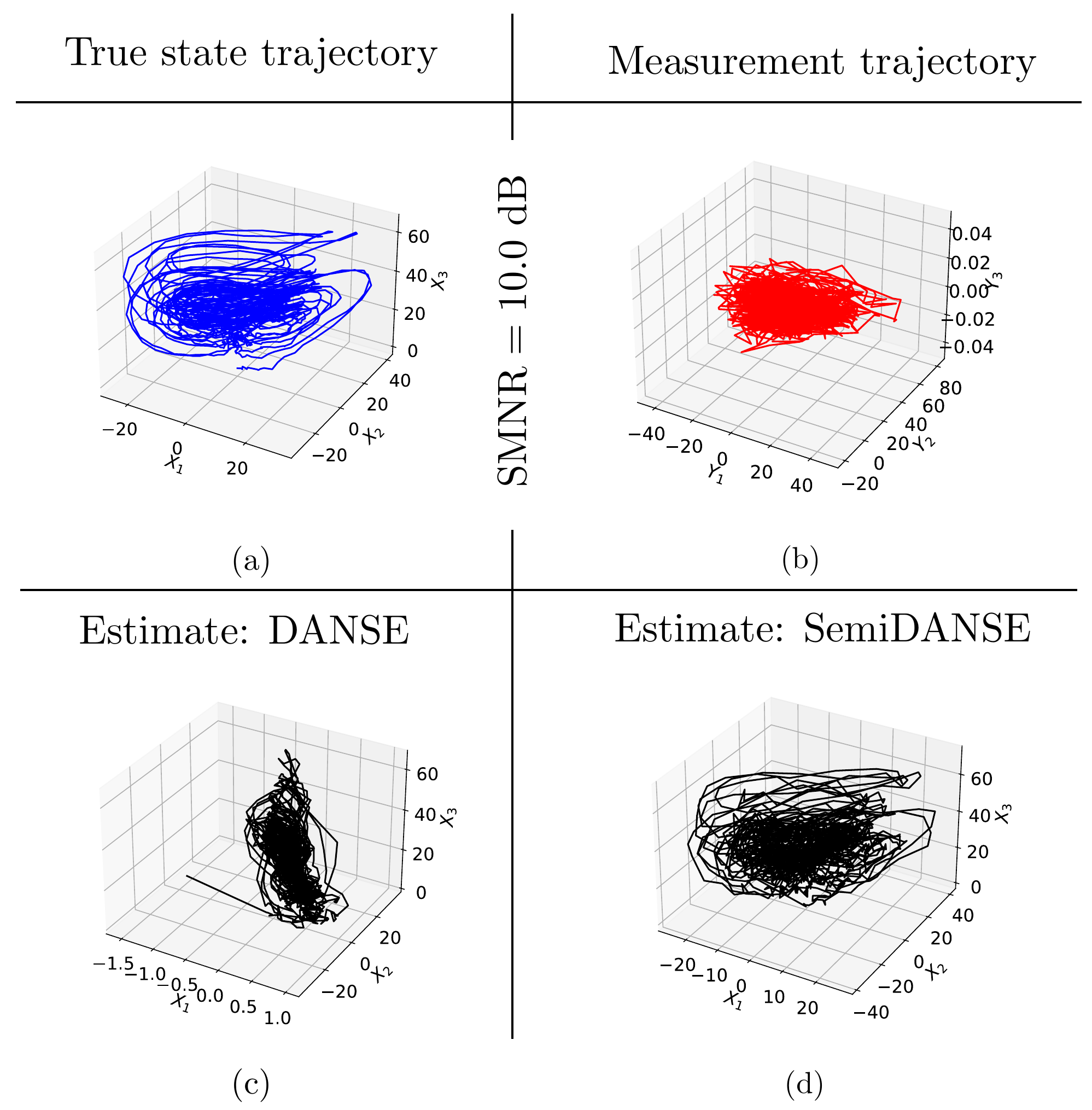}
    \caption{Additional example of the failure of DANSE and success of SemiDANSE for BSCM problem with partial measurement of $3$-dimensional Chen process. We use the $2 \times 3$ measurement matrix $\mathbf{H}$ shown in \eqref{eq:Partial_H}. A $3$-dimensional true state trajectory from $\Dataset_{\text{test}}$ with $\sigmapnoise^2$ corresponding to $-10$ dB is shown in (a). The corresponding measurement trajectory at $\text{SMNR} = 10 \text{ dB}$ is shown in (b). The estimates for DANSE and SemiDANSE ($\fraclabelled=0.02$) are posterior mean estimates shown in (c) and (d) respectively.}
    \label{fig:semidanse_demonstrator_chen}
\end{figure}

\subsection{On predictive performance / forecasting}\label{subsec:predictiveperformance}
In this subsection, we show that DANSE provides a good predictive / forecasting performance for tracking the measurement sequence $\bmeas_t$ despite failing to estimate the underlying state sequence $\bstate_t$, {which} means failing to address the BSCM for the partial measurement system in Section \ref{subsec:PartialMeasurement}. In Fig. \ref{fig:predictiveperformance}, we plot the time-wise and coordinate-wise values of the measurement vector $\bmeas_t$ and the corresponding one-step ahead predicted measurement  $\hat{\bmeas}_t$ given $\bmeas_{1:t-1}$ using DANSE and SemiDANSE at $10$ dB SMNR for the Lorenz-$63$ process. Here $\hat{\bmeas}_t$ is the predictive mean of $p(\bmeas_t|\bmeas_{1:t-1})$ (see \eqref{eq:pyt_given_prev} for SemiDANSE.). A notable fact is that DANSE provides reasonably good predictive performance compared to the proposed SemiDANSE, which shows that the DANSE is able to capture the underlying dynamics of the measurements, but the structure of $\bmeasmat$ in \eqref{eq:Partial_H} renders it difficult to track the unobserved state vector coordinate $x_{t,1}$ as shown in Fig. \ref{fig:axeswise_plot_LorenzSSMn2}. 

This result for DANSE concludes a fact: unsupervised learning may provide a good predictive / forecasting performance on measurement data as a consequence of efficient time-series modelling, but that may not translate to a good state estimation performance. Estimating the hidden state from measurements proves to be a harder task than modeling measurement data.

\subsection{Partial measurement scenario - an extreme BSCM case}
\label{subsec:PartialMeasurementExtreme}
In the previous subsection, we had measurements from two components of the three-dimensional state vector of the Lorenz-$63$ process. An extreme BSCM problem case is: we have only measurements from one state component. For example, we have the following measurement matrix
\begin{eqnarray}\label{eq:Hn1}
    \bmeasmat= \begin{bmatrix}
        1 &  0 & 0 
    \end{bmatrix}.
\end{eqnarray}
That means we have the measurement of the first component of the $3$-dimensional state vector. In this extreme case, Fig. \ref{fig:semidanse_demonstrator_n1} shows a $3$-dimensional trajectory of the Lorenz-$63$ process, and the $3$-dimensional state estimates from DANSE, UKF and SemiDANSE at SMNR $= 10$ dB. We observe again that data-driven SemiDANSE and model-driven UKF perform well, but unsupervised learning-based DANSE fails. {In this regard, we remark that the success of SemiDANSE is because SemiDANSE can exploit the time-wise correlation in the state trajectories in $\Dataset_{\text{semi}}$ and the dimension-wise correlation in each state vector, e.g., in the case of Lorenz-$63$ system, the second and third  components of $\bstate_t$ depend on the first component. 
}

\subsection{{Demonstration for more chaotic dynamical systems}}
\label{sec:experimental_results_demonstration_additional}
{We} also show the performance of SemiDANSE on two other {$3$-dimensional} chaotic dynamical systems - the Chen attractor and the R\"ossler attractor - for partial measurement setups. We use the $2 \times 3$-dimensional $\bmeasmat$ matrix shown in \eqref{eq:Partial_H} and $\text{SMNR}=10 \text{ dB}$. From Fig. \ref{fig:semidanse_demonstrator_chen} and Fig. \ref{fig:semidanse_demonstrator_rossler}, we observe that while DANSE fails, SemiDANSE succeeds. 

\begin{figure}[t]
    \centering
    \includegraphics[width=0.5\textwidth]{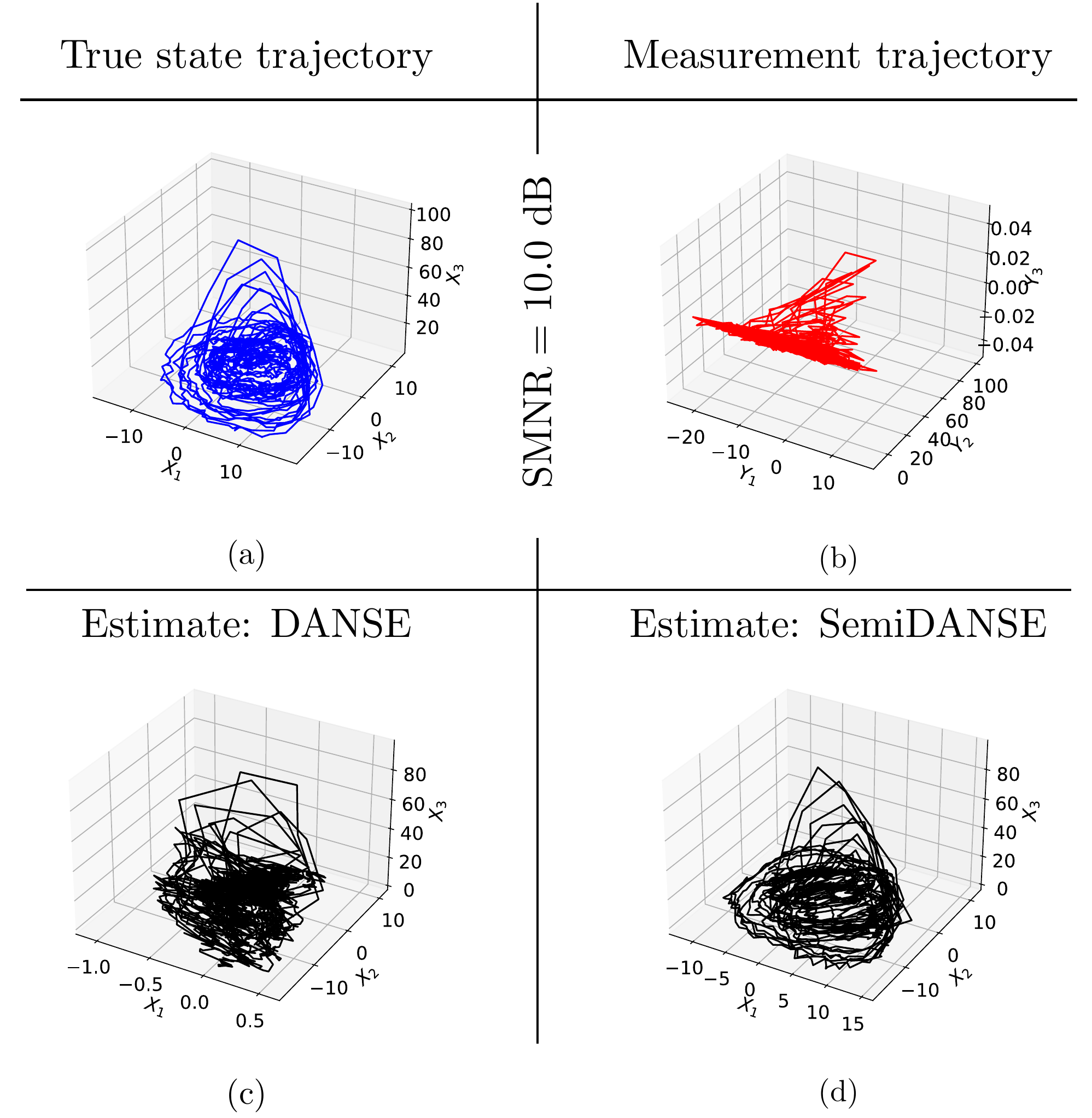}
    \caption{Additional example of the failure of DANSE and success of SemiDANSE for BSCM problem with partial measurement of $3$-dimensional R\"ossler process. We use the $2 \times 3$ measurement matrix $\mathbf{H}$ shown in \eqref{eq:Partial_H}. A $3$-dimensional true state trajectory from $\Dataset_{\text{test}}$ with $\sigmapnoise^2$ corresponding to $-15$ dB is shown in (a). The corresponding measurement trajectory at $\text{SMNR} = 10 \text{ dB}$ is shown in (b). The estimates for DANSE and SemiDANSE ($\fraclabelled=0.02$) are posterior mean estimates shown in (c) and (d), respectively.}
    \label{fig:semidanse_demonstrator_rossler}
\end{figure}
{\begin{figure}[ht]
    \centering
    \scalebox{1.0}{
\begin{tikzpicture}

\definecolor{darkgray176}{RGB}{176,176,176}
\definecolor{darkturquoise0191191}{RGB}{0,191,191}
\definecolor{darkviolet1910191}{RGB}{191,0,191}
\definecolor{goldenrod1911910}{RGB}{191,191,0}
\definecolor{green01270}{RGB}{0,127,0}
\definecolor{lightgray204}{RGB}{204,204,204}

\begin{axis}[
legend cell align={left},
legend style={
  fill opacity=0.8,
  draw opacity=1,
  text opacity=1,
  at={(0.03,0.03)},
  anchor=south west,
  draw=lightgray204
},
tick align=outside,
tick pos=left,
x grid style={darkgray176},
xlabel={$n$},
xmajorgrids,
xminorgrids,
xmin=1.1, xmax=15.5,
xtick style={color=black},
y grid style={darkgray176},
ylabel={NMSE (in dB)},
ymajorgrids,
yminorgrids,
ymin=-10.6086895283312, ymax=0.586061243340373,
ytick style={color=black},
height=0.35\textwidth,
width=0.49\textwidth
]

\path [draw=blue, semithick]
(axis cs:15,-5.3757354170084)
--(axis cs:15,-4.97313104569912);

\path [draw=blue, semithick]
(axis cs:10,-2.7639574855566)
--(axis cs:10,-2.52687503397465);

\path [draw=blue, semithick]
(axis cs:5,-0.705633044242859)
--(axis cs:5,-0.568243145942688);

\path [draw=blue, semithick]
(axis cs:2,-0.270513691008091)
--(axis cs:2,-0.195518337190151);

\path [draw=darkviolet1910191, semithick]
(axis cs:15,-7.2191666662693)
--(axis cs:15,-6.82589539885521);

\path [draw=darkviolet1910191, semithick]
(axis cs:10,-4.82525236904621)
--(axis cs:10,-4.53617112338543);

\path [draw=darkviolet1910191, semithick]
(axis cs:5,-3.09555084258318)
--(axis cs:5,-2.89789932221174);

\path [draw=darkviolet1910191, semithick]
(axis cs:2,-2.1491008028388)
--(axis cs:2,-2.03987357765436);
\addplot [semithick, blue, mark=triangle*, mark size=3, mark options={solid}]
table {%
15 -5.17443323135376
10 -2.64541625976562
5 -0.636938095092773
2 -0.233016014099121
};
\addlegendentry{DANSE}
\addplot [semithick, darkviolet1910191, mark=pentagon, mark size=3, mark options={solid}]
table {%
15 -7.02253103256226
10 -4.68071174621582
5 -2.99672508239746
2 -2.09448719024658
};
\addlegendentry{{SemiDANSE}}

\end{axis}

\end{tikzpicture}}
    \caption{{The average NMSE (in dB) on $\Dataset_{\text{test}}$ versus $n$ performances to illustrate the success of SemiDANSE ($\fraclabelled=0.04, 0.08$) for the BSCM problem setup for the $20$-dimensional  Lorenz-$96$ system (described in Section \ref{sec:experimental_results_demonstration_additional}), {with $\sigma_{\pnoise}^2$ corresponding to $-10$ dB and SMNR corresponding to $10$ dB}. $n$ is varied from $15, 10, 5, 2$ with $m=20$ kept fixed. The comparison is against the unsupervised learning-based DANSE.}}
    \label{fig:nmse_vs_smnr_partial_lorenz96}
\end{figure}
}
{
Additionally, we performed experiments on a $20$-dimensional, stochastic Lorenz-$96$ system \cite{lorenz1996predictability, lorenz1998optimal}. The system details can be found in \cite[Appendix B]{ghosh2023dansejrnl}, and we followed a similar experimental setup as described in \cite[Section H]{ghosh2023dansejrnl}. We investigated the quantitative performance of DANSE and SemiDANSE for the measurement system in \eqref{eq:measurement}, with different sizes of the measurement matrix, starting with $\bmeasmat = \begin{bmatrix}
\mathbf{I}_{15 \times 15} & \mathbf{O}_{15 \times 5} \\
\mathbf{O}_{5 \times 15} & \mathbf{O}_{5 \times 5} \\
\end{bmatrix}$ (corresponds to a partial measurement scenario with $n = 15, m = 20$) by varying $\measdim$, keeping $\statedim$ fixed at $20$. We used $\sigmapnoise^{2}$ corresponding to $-10$ dB and the SMNR corresponding to $10$ dB. Thereafter, we reduced the measurement dimension $n$ to $10, 5, 2$, and in each case, evaluated the performance for both DANSE and SemiDANSE. DANSE was trained using $\Dataset_{u}$ with $N=250, T=1000$, SemiDANSE was trained using $\Dataset_{\text{semi}}$ with $\fraclabelled=0.08$ with the same $N, T$ as in the case of DANSE. Both the methods were tested on the same $\Dataset_{\text{test}}$ with $N_{\text{test}} = 100, T_{\text{test}} = 2000$. The natural expectation is that the difficulty of the BSCM problem increases with the decreasing value of $\measdim$. The results are shown in Fig. \ref{fig:nmse_vs_smnr_partial_lorenz96}. We observe that SemiDANSE achieves a reasonable performance gain over DANSE, which becomes noticeable as $n$ decreases, thus demonstrating the applicability of SemiDANSE to high-dimensional state estimation problems. 
}

\section{Conclusions}\label{sec:conclusions}
{We propose SemiDANSE - a semi-supervised data-driven state estimation method for tackling the BSCM problem in a model-free scenario. We empirically demonstrate the limitation of unsupervised learning for addressing the BSCM task in the case of two notable methods - DANSE and DMM. While DANSE provides satisfactory forecasting performance, the state estimation performance for the BSCM problem is unsatisfactory. Then, we empirically show the success of SemiDANSE in tackling the BSCM problem {using a handful of different measurement systems and benchmark chaotic dynamical systems. We show this by employing only a limited amount of labelled data - for instance, $2\%$ of the total number of training data samples together with unlabelled data for training SemiDANSE.} Compared to DANSE, SemiDANSE shows noticeable improvement in learning, owing to the desired regularization brought about by the limited amount of labelled data. The performance of SemiDANSE also improves with the increased availability of supervision. Future work includes further exploration of the effect of semi-supervised learning for the general case of nonlinear, non-Gaussian measurements. 
}

\section{Acknowledgement}
{The authors would like to thank Dr. Sara Saeidian, KTH, for helpful comments on the structure and flow of the manuscript. 
}
\appendix
\subsection{{Parameterization of the Gaussian prior using RNN}}\label{sec:parameterizationGaussianPriorRNN}
In this subsection, we describe the parameterization of the Gaussian prior using RNN in \eqref{eq:prior_lik}. The schematic representation is shown in Fig. \ref{fig:rnnGaussianprior}. From Fig. \ref{fig:rnnGaussianprior}, {we} note that the prior parameters $\stateMeanprior{t}$ and $\stateCovprior{t}$ are calculated using an RNN together with feed-forward networks using $\bmeas_{1:t-1}$ as input. 

Let the internal state of the RNN at time $\tau$ be $\rnnhstate_\tau \in \mathbb{R}^{\rnnhstatedim}$. For a causal system, we use unidirectional RNNs, where $\rnnhstate_\tau$ generally depends on the \textit{previous} internal state $\rnnhstate_{\tau-1}$ and the current input $\bmeas_{\tau}$ as
\begin{eqnarray}
\begin{array}{rl}
     \rnnhstate_{\tau} = \pmb{\varphi}\left(\mathbf{W}_{\rnnhstateindex \rnnhstateindex} \rnnhstate_{\tau-1} + \mathbf{b}_{\rnnhstateindex \rnnhstateindex} + \mathbf{W}_{\rnnhstateindex \meas} \bmeas_{\tau} + \mathbf{b}_{\rnnhstateindex \meas}\right), 
\end{array}
\end{eqnarray}
where $\mathbf{W}_{\rnnhstateindex \rnnhstateindex} \in \mathbb{R}^{\rnnhstatedim \times \rnnhstatedim}, \mathbf{b}_{\rnnhstateindex \rnnhstateindex} \in \mathbb{R}^{\rnnhstatedim \times 1}, \mathbf{W}_{\rnnhstateindex \meas} \in \mathbb{R}^{\rnnhstatedim \times \statedim}, \mathbf{b}_{\rnnhstateindex \meas} \in \mathbb{R}^{\rnnhstatedim \times 1}$ denote the learnable weights and biases for the connections between the hidden states and the inputs {and} $\pmb{\varphi}\left(\cdot\right)$ denotes a nonlinear function \cite[Chap. 10]{goodfellow2016deep}. For the simplest RNN architectures such as a vanilla RNN, $\pmb{\varphi}\left(\cdot\right)$ can be as simple as an element-wise activation like a hyperbolic tangent function ($\text{tanh}(\cdot)$) or can be much more complicated for sophisticated architectures such as GRUs and LSTMs involving gating functions \cite{choPropertiesNeuralMachine2014, hochreiter1997long}. In our case, $\pmb{\varphi}(\cdot)$ is the nonlinear function for a GRU \cite{choPropertiesNeuralMachine2014}. The parameters of the RNN $\mathbf{W}_{\rnnhstateindex \rnnhstateindex}, \mathbf{b}_{\rnnhstateindex \rnnhstateindex}, \mathbf{W}_{\rnnhstateindex \meas}, \mathbf{b}_{\rnnhstateindex \meas}$ are shared across time and learning happens in accordance with backpropagation through time (BPTT) \cite{werbos1990backpropagation}. 
The internal state at the $({t-1})$'th time instant $\rnnhstate_{t-1}$ captures the information of the sequence $\bmeas_{1:t-1}$. We then map $\rnnhstate_{t-1}$ to the mean $\stateMeanprior{t}$ and the covariance $\stateCovprior{t}$ using two shallow feed-forward networks $\text{FC}_{\stateMean}, \text{FC}_{\stateCov}$ through an intermediate representation $\rnnhstate_{t-1}^{'}$ as follows
\begin{eqnarray}\label{eq:meancovpriorfeedforward}
\begin{array}{rl}
     \rnnhstate_{t-1}^{'} &= \text{ReLU}\left(\mathbf{W}_{\rnnhstateindex' \rnnhstateindex} \rnnhstate_{t-1} + \mathbf{b}_{\rnnhstateindex' \rnnhstateindex} \right), \\
     \stateMeanprior{t} &= \text{FC}_{\stateMean}\left(\rnnhstate_{t-1}\right) \\
     &= \mathbf{W}_{m\rnnhstateindex'}\rnnhstate_{t-1}^{'} 
     + \mathbf{b}_{m\rnnhstateindex'},\\
     \stateCovprior{t} &= \text{diag}\left(\text{FC}_{\stateCov}\left(\rnnhstate_{t-1}\right)\right) \\
     &= \text{diag}\left(\text{softplus}\left(\mathbf{W}_{L\rnnhstateindex'} \rnnhstate_{t-1}^{'} + \mathbf{b}_{L\rnnhstateindex'}\right)\right),\\
\end{array}
\end{eqnarray}
where $\mathbf{W}_{m\rnnhstateindex'} \in \setR^{\statedim \times \rnnhstatedim'}, \mathbf{b}_{m\rnnhstateindex'} \in \setR^{\statedim \times 1}, \mathbf{W}_{L\rnnhstateindex'} \in \setR^{\statedim \times \rnnhstatedim'}, \mathbf{b}_{L\rnnhstateindex'} \in \setR^{\statedim \times 1}, \mathbf{W}_{\rnnhstateindex' \rnnhstateindex} \in \setR^{\rnnhstatedim' \times \rnnhstatedim}, \mathbf{b}_{\rnnhstateindex' \rnnhstateindex} \in \setR^{\rnnhstatedim' \times 1}$ are learnable weights and biases, $\text{ReLU}(x) = \max\left(0, x\right)$ is the element-wise, rectified linear unit function and $\text{softplus}(x) = \log_{e}\left(1 + \exp({x})\right)$ is the element-wise, smooth approximation to $\text{ReLU}(x)$ \cite{paszke2019pytorch}. As explained in section \ref{sec:inferenceproblem}, $\stateCovprior{t}$ is modelled as a diagonal covariance matrix {as per }\eqref{eq:meancovpriorfeedforward}. Thus, as per the notation in \eqref{eq:prior_lik}, the full set of learnable parameters is $\btheta = \lbrace\mathbf{W}_{\rnnhstateindex \rnnhstateindex}, \mathbf{b}_{\rnnhstateindex \rnnhstateindex}, \mathbf{W}_{\rnnhstateindex \meas}, \mathbf{b}_{\rnnhstateindex \meas},  \mathbf{W}_{m\rnnhstateindex'}, \mathbf{b}_{m\rnnhstateindex'}, \mathbf{W}_{L\rnnhstateindex'}, \mathbf{b}_{L\rnnhstateindex'}, \mathbf{W}_{\rnnhstateindex' \rnnhstateindex}, \mathbf{b}_{\rnnhstateindex' \rnnhstateindex} \rbrace$. The specific RNN architectures and the feed-forward networks are chosen by grid-search. and described in Section \ref{sec:trainingtestinghyperparams}. 
\subsection{Three chaotic dynamical systems}\label{sec:chaoticdynsystems_detail}
In this subsection, we briefly describe the mathematical models of the three chaotic dynamical systems used for experiments in section \ref{sec:chaoticdynsystems}. The first two chaotic dynamical systems - the Lorenz-$63$ attractor \cite{lorenz1963deterministic} and the Chen attractor \cite{chen1999yet} were developed independently but have been shown to possess a generalized Lorenz canonical form in \cite{vcelikovsky2005generalized}. We used the discretized form of the Lorenz-$63$ attractor described in \cite{ghosh2023dansejrnl}, as follows:
\begin{eqnarray}\label{eq:lorenz_state}
\begin{array}{rl}
     \bstate_{t+1} &= \bfn_{t}^{\text{(L)}}(\bstate_{t}) + \bpnoise_t \in \setR^3, \\
     &= \mathbf{F}_{t}^{\text{(L)}}(\bstate_{t})\bstate_{t} + \bpnoise_t \in \setR^3, \\
\end{array}
\end{eqnarray}
\begin{eqnarray*}\label{eq:lorenz_mexp}
\begin{array}{rl}
    \text{s.t. }\mathbf{F}_{t}^{\text{(L)}}(\bstate_{t}) &= \matrixexp{\left(
    \begin{bmatrix}
        -10 & 10 & 0 \\
        28 & -1 & -\state_{t, 1} \\
        0 & \state_{t, 1} & -\dfrac{8}{3} \\ 
    \end{bmatrix}\Delta 
    \right)}, \\
\end{array}   
\end{eqnarray*}
where the process noise $\bpnoise_t \sim \normaldist{\bpnoise_t}{\boldsymbol{0}}{\bpnoiseCov}$ with $\bpnoiseCov=\sigmapnoise^2 \mathbf{I}_{3}$, the step-size $\Delta=0.02 \text{ seconds}$. $\sigmapnoise^2$ was set corresponding to $-10$ dB. The subscript $(L)$ refers to the Lorenz attractor. In our simulations, we use a finite-Taylor series approximation of $5'\text{th}$ order for $\mathbf{F}_{t}^{\text{(L)}}(\bstate_{t})$. 

The Chen attractor has a mathematically similar form to the Lorenz-$63$ attractor except that the constants in the {dynamical model} are different. Similar to \eqref{eq:lorenz_state}, it has the following mathematical form 
\begin{eqnarray}\label{eq:chen_state}
\begin{array}{rl}
     \bstate_{t+1} 
     &= \mathbf{F}_{t}^{\text{(C)}}(\bstate_{t})\bstate_{t} + \bpnoise^{\text{(C)}}_t \in \setR^3, \\
\end{array}
\end{eqnarray}
\begin{eqnarray*}\label{eq:chen_mexp}
\begin{array}{rl}
    \text{s.t. }\mathbf{F}_{t}^{\text{(C)}}(\bstate_{t}) &= \matrixexp{\left(
    \begin{bmatrix}
        -35 & 35 & 0 \\
        -7 & 28 & -\state_{t, 1} \\
        0 & \state_{t, 1} & -{3} \\ 
    \end{bmatrix}\Delta^{\prime} 
    \right)}, \\
\end{array}   
\end{eqnarray*}
where $\bpnoise^{\text{(C)}}_t \sim \normaldist{\bpnoise^{\text{(C)}}_t}{\boldsymbol{0}}{\bpnoiseCov^{\text{(C)}}}$ with $\bpnoiseCov^{\text{(C)}}=\sigmapnoise^2 \mathbf{I}_{3}$, the step-size $\Delta^{\prime}=0.002 \text{ seconds}$. $\sigmapnoise^2$ was set corresponding to $-10$ dB. In our simulations, we use a finite-Taylor series approximation of $5$'$\text{th}$ order for $\mathbf{F}_{t}^{\text{(C)}}(\bstate_{t})$.

The R\"ossler attractor \cite{rossler1976equation} appears simpler than the above two in terms of non-linear relationships among the state variables. Similar to \eqref{eq:lorenz_state}, \eqref{eq:chen_state}, the mathematical form is
\begin{eqnarray}\label{eq:rossler_state}
\begin{array}{rl}
     \bstate_{t+1} 
     &= \mathbf{F}_{t}^{\text{(R)}}(\bstate_{t})\bstate_{t} + \bpnoise^{\text{(R)}}_t \in \setR^3, \\
\end{array}
\end{eqnarray}
\begin{eqnarray*}\label{eq:rossler_mexp}
\begin{array}{rl}
    &\text{s.t. }\mathbf{F}_{t}^{\text{(R)}}(\bstate_{t}) = \matrixexp{\left(
    \begin{bmatrix}
        0 & -1 & -1 \\
        1 & 0.2 & 0 \\
        0 & 0 & \frac{0.2}{\state_{t,3}} + (\state_{t,1} - 5.7) \\ 
    \end{bmatrix}\Delta^{\prime\prime} 
    \right)}, \\
\end{array}   
\end{eqnarray*}
where $\bpnoise^{\text{(R)}}_t \sim \normaldist{\bpnoise^{\text{(R)}}_t}{\boldsymbol{0}}{\bpnoiseCov^{\text{(R)}}}$ with $\bpnoiseCov^{\text{(R)}}=\begin{bmatrix}
 \sigmapnoise^2 \mathbf{I}_{2}  & \boldsymbol{0}_{2\times1} \\
 \boldsymbol{0}_{1\times2}  & \epsilon \\
\end{bmatrix}$, 
$\epsilon \in \setR_{+}$ is constant, the step-size $\Delta^{\prime\prime}=0.008 \text{ seconds}$. In our simulations, we use a finite-Taylor series approximation of $5$'$\text{th}$ order for $\mathbf{F}_{t}^{\text{(R)}}(\bstate_{t})$. For all the three systems, the measurement system is the linear measurement system in \eqref{eq:measurement}, with the measurement noise is $\bmnoise_t\sim \normaldist{\bmnoise_t}{\boldsymbol{0}}{\bmnoiseCov}$ with $\bmnoiseCov=\sigmamnoise^2 \mathbf{I}_{3}$. $\sigmapnoise^2$ was set corresponding to $-15               $ dB. The truncated, reduced process noise variance is used in the simulation of the R\"ossler attractor in \eqref{eq:rossler_state} to avoid numerical problems owing to incorporating noise in the dynamics. Since all the systems do not have the same step-size, in order to get a trajectory of length $T$ with a similar degree of captured dynamics as the Lorenz-$63$ attractor, the other ones viz. Chen and R\"ossler are simulated for longer $T$, and then decimated in time by an appropriate factor (depending on the step-size) to get a {$T$-length trajectory}.
\subsection{DMM - Learning problem}\label{sec:dmmdetail}
The deep Markov model (DMM) method proposed in \cite{krishnan2017structured} is well-suited to the modeling of complex time-series signals. A detailed description of the different inference modes and training architectures of the DMM can be found in \cite{krishnan2017structured}, \cite[Chap. 5]{girin2021dynamical}. Here, we provide a brief description of the DMM that we used for comparison in section \ref{sec:experimentsandresults}.
The DMM method relies on a choice of an \textit{approximate} posterior distribution $q\left(\bstate_{1:T} \vert \bmeas_{1:T}; \pmb{\phi} \right)$ parameterized using a deep recurrent neural network with parameters $\pmb{\phi}$. The factorization of this approximate posterior defines the inference mode used. E.g. in this work, we employed the structured left-information (ST-L) mode, as it gives rise to a causal filtering scenario \cite{krishnan2017structured}. The approximate posterior would then factorize as 
\begin{eqnarray}\label{eq:st_l_q}
\begin{array}{rl}
     q\left(\bstate_{1:T} \vert \bmeas_{1:T}; \pmb{\phi} \right) = \prod\limits_{t=1}^T q\left(\bstate_t \vert \bstate_{t-1} \bmeas_{1:t} ; \pmb{\phi}\right).
\end{array}    
\end{eqnarray}
In addition to this, the DMM method assumes that the process state dynamics is Markovian and tries to learn a model  $p\left(\bstate_{t} \vert \bstate_{t-1}; \pmb{\psi} \right)$ with learnable parameters $\pmb{\psi}$. This is achieved by optimizing a variational lower bound (VLB) using \eqref{eq:measurement}, \eqref{eq:st_l_q} 
and the Markovian assumption as follows:
\begin{eqnarray*}\label{eq:vlb_STL}
\begin{array}{lr}
     \log p\left(\bmeas_{1:T}; \pmb{\psi}\right) \\
     = \log \left(\bigintsss \cdots \bigintsss  p\left(\bmeas_{1:T}, \bstate_{1:T}; \pmb{\psi}\right) d\bstate_{1} \cdots d\bstate_{T}\right)  \\
     = \log \left(\mathbb{E}_{q\left(\bstate_{1:T} \vert \bmeas_{1:T}; \pmb{\phi} \right)}\left[{\dfrac{p\left(\bmeas_{1:T}, \bstate_{1:T}; \pmb{\psi}\right)}{q\left(\bstate_{1:T} \vert \bmeas_{1:T}; \pmb{\phi} \right)}}\right]\right)  \\
     \mygeq{(a4)}
     \mathbb{E}_{q\left(\bstate_{1:T} \vert \bmeas_{1:T}; \pmb{\phi} \right)} \left[\log \left(\dfrac{p\left(\bmeas_{1:T} \vert \bstate_{1:T}; \pmb{\psi}\right)p\left(\bstate_{1:T}; \pmb{\psi}\right)}{q\left(\bstate_{1:T} \vert \bmeas_{1:T}; \pmb{\phi} \right)}  \right)\right] \\
     =\mathbb{E}_{q\left(\bstate_{1:T} \vert \bmeas_{1:T}; \pmb{\phi} \right)} \left[\log p\left(\bmeas_{1:T} \vert \bstate_{1:T}; \pmb{\psi} \right)\right] \\
     + \mathbb{E}_{q\left(\bstate_{1:T} \vert \bmeas_{1:T}; \pmb{\phi} \right)} \left[\log \left(\dfrac{p\left(\bstate_{1:T}; \pmb{\psi}\right)}{q\left(\bstate_{1:T} \vert \bmeas_{1:T}; \pmb{\phi} \right)}  \right)\right] \\
     \myeq{(b4)} \sum_{t=1}^T \mathbb{E}_{q \left(\bstate_{t} \vert \bmeas_{1:t}; \pmb{\phi} \right)} \left[\log \mathcal{N}(\bmeas_t ; \bmeasmat\bstate_t, \mathbf{C}_w )\right] \\
     - \mathbb{E}_{q \left(\bstate_{t-1} \vert \bmeas_{1:t-1}; \pmb{\phi} \right)} \left[\text{D}_{\text{KL}}\left( q \left(\bstate_{t} \vert \bstate_{t-1}, \bmeas_{1:t}; \pmb{\phi} \right) \Vert p\left(\bstate_{t} \vert \bstate_{t-1} ; \pmb{\psi}\right) \right)\right] \\
     \triangleq \mathcal{L}\left( \bmeas_{1:T}; \pmb{\psi, \phi}\right),
\end{array}
\end{eqnarray*}
where the inequality in (a4) is obtained by Jensen's inequality and the detailed derivation of the equality in (b4) is detailed in \cite[Chap. 5]{girin2021dynamical}. In practice, we evaluate the above expectations using Monte-Carlo approximations and sequential sampling with the reparameterization trick \cite{kingma2014auto}, making it quite computationally expensive. For a training dataset $\Dataset = \lbrace \bmeas^{(i)}_{1:T^{(i)}} \rbrace_{i=1}^N$, the optimization problem then becomes 
\begin{eqnarray}\label{eq:modified_optimprob}
\begin{array}{rl}
     \pmb{\psi}^{\star}, \pmb{\phi}^{\star} 
     &= \argmin\limits_{\pmb{\psi}, \pmb{\phi}} \sum_{i=1}^N -\mathcal{L}\left(\bmeas^{(i)}_{1:T^{(i)}}; ; \pmb{\psi, \phi}\right). \\
\end{array}
\end{eqnarray}
The inference follows by sampling sequentially from the learned, approximate posterior $q\left(\bstate_t \vert \bstate_{t-1} \bmeas_{1:t}; \pmb{\phi}^{\star}\right)$ using the reparameterization trick. 
\bibliographystyle{IEEEbib}
\bibliography{strings,refs}
\end{document}